\documentclass[
 reprint,
nofootinbib,
 amsmath,amssymb,
 aps,
floatfix %I've turned this on in response to error messages -CTB
]{revtex4-1}

\usepackage{graphicx}% Include figure files
\usepackage{dcolumn}% Align table columns on decimal point
\usepackage{bm}% bold math
\usepackage[colorinlistoftodos]{todonotes}
\usepackage[export]{adjustbox}%
\usepackage{natbib}
\newcommand{\aact}{\ensuremath{\alpha_{\textrm{act}}}}
\newcommand{\anom}{\ensuremath{\alpha_{\textrm{nom}}}}

\begin{document}

\title{Publication bias and the canonization of false facts}

\thanks{KG and CTB contributed equally and share the senior author role.}

\author{Silas B. Nissen}
\email{silas@nbi.ku.dk}
\affiliation{Niels Bohr Institute \\ University of Copenhagen \\ Copenhagen, Denmark}

\author{Tali Magidson}
\email{talim@uw.edu}
\affiliation{Department of Computer Science \\ University of Washington \\ Seattle, WA USA}

\author{Kevin Gross}
\email{krgross@ncsu.edu}
\affiliation{Department of Statistics \\ North Carolina State University \\ Raleigh, NC USA}

\author{Carl T. Bergstrom}
\email{cbergst@u.washington.edu}
\affiliation{Department of Biology \\ University of Washington \\ Seattle, WA USA}

\date{\today}

\begin{abstract}

Science is facing a ``replication crisis" in which many experimental findings are irreplicable and likely false. Does this imply that many scientific facts are false as well? To find out, we explore the process by which a claim becomes fact. We model the community's confidence in a claim as a Markov process with successive published results shifting the degree of belief. Publication bias in favor of positive findings influences the distribution of published results. We find that unless a sufficient fraction of negative results are published, false claims frequently can become canonized as fact. Data-dredging, p-hacking, and similar behaviors exacerbate the problem. When results become easier to publish as a claim approaches acceptance as a fact, however, true and false claims can be more readily distinguished. To the degree that the model reflects the real world, there may be serious concerns about the validity of purported facts in some disciplines.

\end{abstract}

\maketitle

\section{Introduction}

Science is a process of collective knowledge creation in which researchers use experimental, theoretical, and observational approaches to develop a naturalistic understanding of the world. In the development of a scientific field, certain claims stand out as both significant and stable in the face of further experimentation \cite{ravetz1971scientific}. Once a claim reaches this stage of widespread acceptance as true, it has transitioned from claim to {\em fact}. This transition, which we call {\em canonization}, is often indicated by some or all of the following: a canonized fact can be taken for granted rather than treated as an open hypothesis in the subsequent primary literature; tests that do no more than to confirm previously canonized facts are seldom considered publication-worthy; and canonized facts begin to appear in review papers and textbooks without the company of alternative hypotheses. Of course the veracity of so-called facts may be called back into question \cite{arbesman2012halflife,latour1987science}, but for time being the issue is considered to be settled. Note that we consider facts to be epistemological rather than ontological: a claim is a fact because it is accepted by the relevant community, not because it accurately reflects or represents underlying physical reality \cite{ravetz1971scientific,latour1987science}.

But what is the status of these facts in light of the replication crisis purportedly plaguing science today? Large scale analyses have revealed that many published papers in fields ranging from cancer biology to psychology to economics cannot be replicated in subsequent experiments \cite{begley2012drug,open2015estimating,errington2014open,ebrahim2014reanalyses,chang2015economics,camerer2016evaluating,Baker2016}. One possible explanation is that many published experiments are not replicable because many of their conclusions are ontologically false \cite{Ioannidis2005,higginson2016current}.

If many experimental findings are ontologically false, does it follow that many scientific facts are ontologically untrue? Not necessarily. Claims of the sort that become facts are rarely if ever tested directly in their entirety. Instead, such claims typically comprise multiple subsidiary hypotheses which must be individually verified. Thus multiple experiments are usually required to establish a claim. Some of these may include direct replications, but more typically an ensemble of distinct experiments will produce multiple lines of evidence before a claim is accepted by the community.

For example, as molecular biologists worked to unravel the details of the eukaryotic RNA interference (RNAi) pathway in the early 2000s, they wanted to understand how the RNAi pathway was initiated. Based on work with {\em Drosophila} cell lines and embryo extracts, one group of researchers made the claim that the RNAi pathway is initiated by the Dicer enzyme which slices double-stranded RNA into short fragments of 20-22 amino acids in length \cite{bernstein2001role}. Like many scientific facts, this claim was too broad to be validated directly in a single experiment. Rather, it comprised a number of subsidiary assertions: an enzyme called Dicer exists in eukaryotic cells; it is essential to initiate the RNAi pathway; it binds dsRNA and slices it into pieces; it is distinct from the enzyme or enzyme complex that destroys targeted messenger RNA; it is ubiquitous across eukaryotes that exhibit RNAi pathway. Researchers from numerous labs tested these subsidiary hypotheses or aspects thereof to derive numerous lines of convergent evidence in support of the original claim. While the initial breakthrough came from work in {\em Drosophila melanogaster} cell lines \cite{bernstein2001role}, subsequent research involved in establishing this fact drew upon {\em in vitro} and {\em in vivo} studies, genomic analyses, and even mathematical modeling efforts, and spanned species including the fission yeast {\em Schizosaccharomyces pombe}, the protozoan {\em Giardia intestinalis}, the nemotode {\em Caenorhabditis elegans}, the flowering plant {\em Arabidopsis thaliana}, mice, and humans \cite{jaskiewicz2008role}. Ultimately, sufficient supporting evidence accumulated to establish as fact the original claim about Dicer's function.

Requiring multiple studies to establish a fact is no panacea, however. The same processes that allow publication of a single incorrect result can also lead to the accumulation of sufficiently many incorrect findings to establish a false claim as fact \cite{mcelreath2015replication}.

This risk is exacerbated by {\em publication bias} \cite{sterling1959publication,rosenthal1979file,newcombe1987towards,begg1988publication,dickersin1990existence,easterbrook1991publication,song2000publication,olson2002publication,chan2005identifying,franco2014publication}. Publication bias arises when the probability that a scientific study is published is not independent of its results \cite{sterling1959publication}. As a consequence, the findings from published tests of a claim will differ in a systematic way from the findings of all tests of the same claim \cite{song2000publication,turner2008selective}.

Publication bias is pervasive. Authors have systematic biases regarding which results they consider worth writing up; this is known as the ``file drawer problem'' or ``outcome reporting bias'' \citep{rosenthal1979file,chan2005identifying}. Journals similarly have biases about which results are worth publishing. These two sources of publication bias act equivalently in the model developed here, and thus we will not attempt to separate them. Nor would separating them be simple; even if authors' behavior is the larger contributor to publication bias \cite{olson2002publication,franco2014publication}, they may simply be responding appropriately to incentives imposed by editorial preferences for positive results.

What kinds of results are most valued? Findings of statistically significant differences between groups or treatments tend to be viewed as more worthy of submission and publication than those of non-significant differences. Correlations between variables are often considered more interesting than the absence of correlations. Tests that reject null hypotheses are commonly seen as more noteworthy than tests that fail to do so. Results that are interesting in any of these ways can be described as ``positive''.

A substantial majority of the scientific results published appear to be positive ones \cite{csada1996file}. It is relatively straightforward to measure the fraction of published results that are negative. One extensive study found that in 2007, more than 80\% of papers reported positive findings, and this number exceeded 90\% in disciplines such as psychology and ecology \cite{fanelli2011negative}. Moreover, the fraction of publications reporting positive results has increased over the past few decades. While this high prevalence of positive results could in principle result in part from experimental designs with increasing statistical power and a growing preference for testing claims that are believed likely to be true, publication bias doubtless contributes as well \cite{fanelli2011negative}. 

How sizable is this publication bias? To answer that, we need to estimate the fraction of negative results that are published, and doing so can be difficult because we rarely have access to the set of findings that go unpublished. The best available evidence of this sort comes from registered clinical trials. For example, in 2008 meta-analysis of 74 FDA-registered studies of antidepressants \cite{turner2008selective}. In that analysis, 37 of 38 positive studies were published as positive results, but only 3 of 24 negative studies were published as negative results. An additional 5 negative studies were re-framed as positive for the purposes of publication. Thus, negative studies were published at scarcely more than 10\% the rate of positive studies.

We would like to understand how the possibility of misleading experimental results and the prevalence of publication bias shape the creation of scientific facts. Mathematical models of the scientific process can help us understand the dynamics by which scientific knowledge is produced and, consequently, the likelihood that elements of this knowledge are actually correct. In this paper, we look at the way in which repeated efforts to test a scientific claim establish this claim as fact or cause it to be rejected as false. 

We develop a mathematical model in which successive publications influence the community's perceptions around the likelihood of a given scientific claim. Positive results impel the claim toward fact, whereas negative results lead in the opposite direction. Describing this process, Bruno Latour \cite{latour1987science} compared the fate of a scientific claim to that of a rugby ball, pushed alternatively toward fact or falsehood by the efforts of competing teams, its fate determined by the balance of their collective actions. Put in these terms, our aim in the present paper is to develop a formal model of how the ball is driven up and down the epistemological pitch until one of the goal lines is reached. In the subsequent sections, we outline the model, explain how it can be analyzed, present the results that we obtain, and consider its implications for the functioning of scientific activity.

\section{Model}

In this section, we will develop a simplified model of scientific activity, designed to capture the important qualitative features of fact-creation as a dynamic process. 

\subsection{Model description}

We explore a simple model in which researchers sequentially test a single claim until the scientific community becomes sufficiently certain of its truth or falsehood that no further experimentation is needed. Our model is conceptually related to those developed in refs.~\cite{rzhetsky2006microparadigms,mcelreath2015replication}, though it is considerably simpler than either since we only consider a single claim at a time.

\begin{figure}
\centering
\includegraphics[width=\columnwidth]{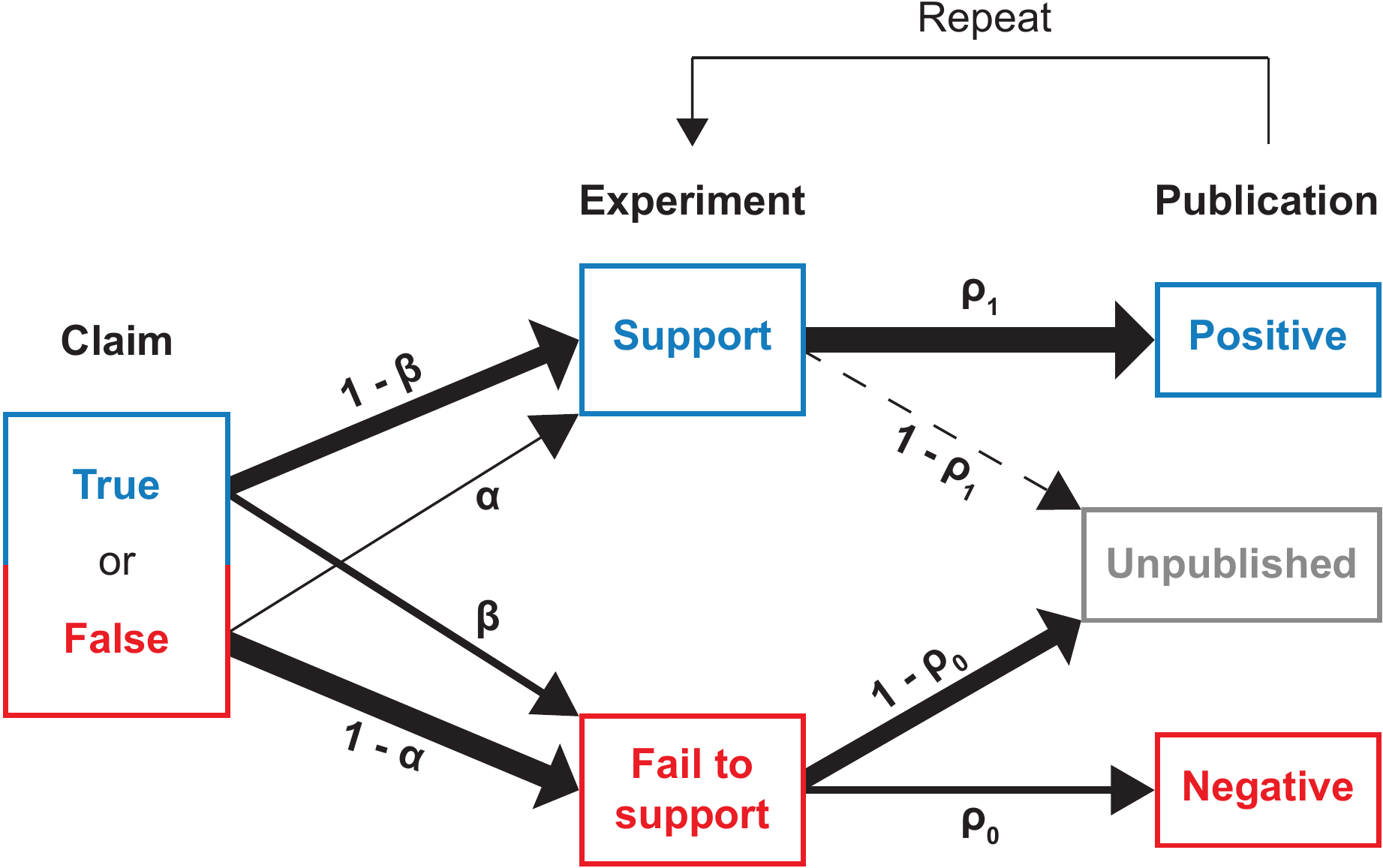}
\caption{{\bf Conducting and reporting the test of a claim}. In our model, a scientific claim is either true or false. Researchers conduct an experiment which either supports or fails to the support the claim. True claims are correctly supported with probability $1-\beta$ while false claims are incorrectly supported with probability $\alpha$. Next, the researchers may attempt to publish their results. Positive results that support the claim are published with probability $\rho_1$ whereas negative results that fail to support the claim are published with probability $\rho_0$. This process then repeats, with additional experiments conducted until the claim is canonized as fact or rejected as false.}
\label{fig:SingleStepSchematic}
\end{figure}

Figure \ref{fig:SingleStepSchematic} provides a schematic illustration of the experimentation and publication process. We begin with a claim which is ontologically either true or false. Researchers sequentially conduct experiments to test the claim; these experiments are typically not direct replications of one another, but rather distinct approaches that lend broader support to the claim. Each experiment returns either a positive outcome supporting the claim, or a negative outcome contravening it. For mathematical simplicity, we assume all tests to have the same error rates, in the sense that if the claim under scrutiny is false, then investigators obtain false positives with probability $\alpha$. Conversely, when the claim is true, investigators obtain false negatives with probability $\beta$. We take these error rates to be the ones that are conventionally associated with statistical hypothesis testing, so that $\alpha$ is equivalent to the significance level (technically, the {\em size}) of a statistical test and $1-\beta$ is the test's power. We assume that, as in any reasonable test, a true claim is more likely to generate a positive result than a negative one: $1-\beta>\alpha$. A broader interpretation of $\alpha$ and $\beta$ beyond statistical error does not change the interpretation of our results. 

After completing a study, the investigators may attempt to publish their experimental results. However, publication bias occurs in that the result of the experiment influences the chance that a study is written up as a paper and accepted for publication. Positive results are eventually published somewhere with probability $\rho_1$ while negative results are eventually published somewhere with probability $\rho_0$. Given the reluctance of authors to submit negative results and of journals to publish them, we expect that in general $\rho_1>\rho_0$.

Finally, readers attempt to judge whether a claim is true by consulting the published literature only. For modeling purposes, we will consider a best-case scenario, in which the false positive and false negative rates $\alpha$ and $\beta$ are established by disciplinary custom or accepted benchmarks, and readers perform Bayesian updating of their beliefs based upon these known values. In practice, these values may not be as well standardized or widely reported as would be desirable. Moreover, readers are unlikely to be this sophisticated in drawing their inferences. Instead readers are likely to form subjective beliefs in an informal fashion based on a general assessment of the accumulated positive and negative results and the strength of each. But the Bayesian updating case provides a well-defined model under which to explore the distortion of belief by publication bias.

The problem is that the results described in the published literature are now biased by the selection of which articles are drafted and accepted for publication. We assume that readers are unaware of the degree of this bias, and that they fail to correct for publication bias in drawing inferences from the published data. It may seem pessimistic that researchers would fail to make this correction, but much of the current concern over the replication crisis in science is predicated on exactly this. Moreover, it is usually impossible for a researcher to accurately estimate the degree of publication bias in a given domain. 

\subsection{Model dynamics}

Consider a claim that the community initially considers to have probability $q_0$ of being true. Researchers iteratively test hypotheses that bear upon the claim until it accumulates either sufficient support to be canonized as fact, or sufficient counter-evidence to be discarded as false. If the claim is true, the probability that a single test leads to a positive publication is $(1-\beta)\rho_1$, and the corresponding probability of a negative publication is $\beta \rho_0$. The remaining probability corresponds to results of the test not being published. If the claim is false, these probabilities are $\alpha\rho_1$ and $(1-\alpha)\rho_0$ for positive and negative published outcomes, respectively. Given that a claim is true, the probability that a published test of that claim reports a positive outcome is therefore 
\begin{equation}
\omega_T=\frac{(1-\beta)\rho_1}{(1-\beta)\rho_1+\beta\rho_0}.
\label{eqn:posiftrue}
\end{equation} 
For a false claim, the probability that a published test is positive is 
\begin{equation}\omega_F=\frac{\alpha \rho_1}{\alpha\rho_1+(1-\alpha)\rho_0}.\label{eqn:posiffalse}
\end{equation} 
Because only the ratio of $\rho_1$ to $\rho_0$ matters for the purposes of our model, we set $\rho_1$ to 1 for the remainder of the paper. We initially assume that $\rho_0$ is constant, but will relax this latter assumption later. 
 
To formalize ideas, consider a sequence of published outcomes $X$, and let $Y_k$ be the number of positive published outcomes in the first $k$ terms of $X$. When the probabilities of publishing a negative result $\rho_0$ is constant, the outcomes of published experiments are exchangeable random variables. Thus after $k$ published tests, the distribution of $Y_k$ for a true claim is the binomial distribution $\mbox{Bin}(k,\omega_T)$ and for a false claim is $\mbox{Bin}(k,\omega_F)$. Moreover, the sequence $\left\{Y_k\right\}_{k=1}^\infty$ is a Markov chain. When the extent of publication bias is known, we can compute the conditional probability that a claim is true, given $Y_k=y$, as
\begin{equation}
\dfrac{\omega_T^y (1-\omega_T)^{k-y}q_0}{\omega_T^y (1-\omega_T)^{k-y}q_0 + \omega_F^y (1-\omega_F)^{k-y}(1-q_0)}.
\label{eq:posterior_belief_correct}
\end{equation}

We now consider the consequences of drawing inferences based on the published data alone, without correcting for publication bias. For model readers who do not condition on publication bias, let $q_k(y)$ be the perceived, conditional probability that a claim is true given $Y_k=y$. We say ``perceived'' because these readers use Bayes' Law to update $q_k$, but do so under the incorrect assumption that there is no publication bias, i.e., that $\rho_0=\rho_1=1$. To ease the narrative, we refer to the perceived conditional probability that a claim is true as the ``belief'' that the claim is true. Expressing this formally, 
\begin{equation}
q_k(y) = \dfrac{(1-\beta)^y \beta^{k-y}q_0}{(1-\beta)^y \beta^{k-y}q_0 + \alpha^y (1-\alpha)^{k-y}(1-q_0)}.
\label{eq:posterior_belief}
\end{equation}
Note that without publication bias, we have $\omega_T = (1-\beta)$ and $\omega_F = \alpha$, and thus eq.~\ref{eq:posterior_belief_correct} coincides with eq.~\ref{eq:posterior_belief}. 

From the perspective of an observer who is unaware of any publication bias, the pair $(Y_k,k)$ is a sufficient statistic for the random variable $A\in\{\mbox{True},\mbox{False}\}$ representing the truth value of the claim in question. This follows from the definition of statistical sufficiency and the fact that 
\begin{eqnarray*}
\mbox{prob} && \left[A=\mbox{True}|Y_k,k,\left\{Y_i\right\}_{i=1}^k\right] = q_k(y) \\ && =\mbox{prob}\left[A=\mbox{True}|Y_k,k\right].
\end{eqnarray*}
By analogous logic, the pair $(Y_k,k)$ is also a sufficient statistic for an observer aware of the degree of publication bias provided that the publication probabilities $\rho_0$ and $\rho_1$ are constant.

We envision science as proceeding iteratively until the belief that a claim is true is sufficiently close to 1 that the claim is canonized as fact, or until belief is sufficiently close to 0 that the claim is discarded as false. We let $\tau_0$ and $\tau_1$ be the belief thresholds at which a claim is rejected or canonized as fact, respectively ($0 < \tau_0 < \tau_1 < 1$), and refer to these as evidentiary standards. In our analysis, we make the simplifying assumption that the evidentiary standards are symmetric, i.e., $\tau_0 = 1 - \tau_1$. We describe the consequences of relaxing this assumption in the Discussion.

\begin{figure*}
\centering
\includegraphics[width=3in,valign=t]{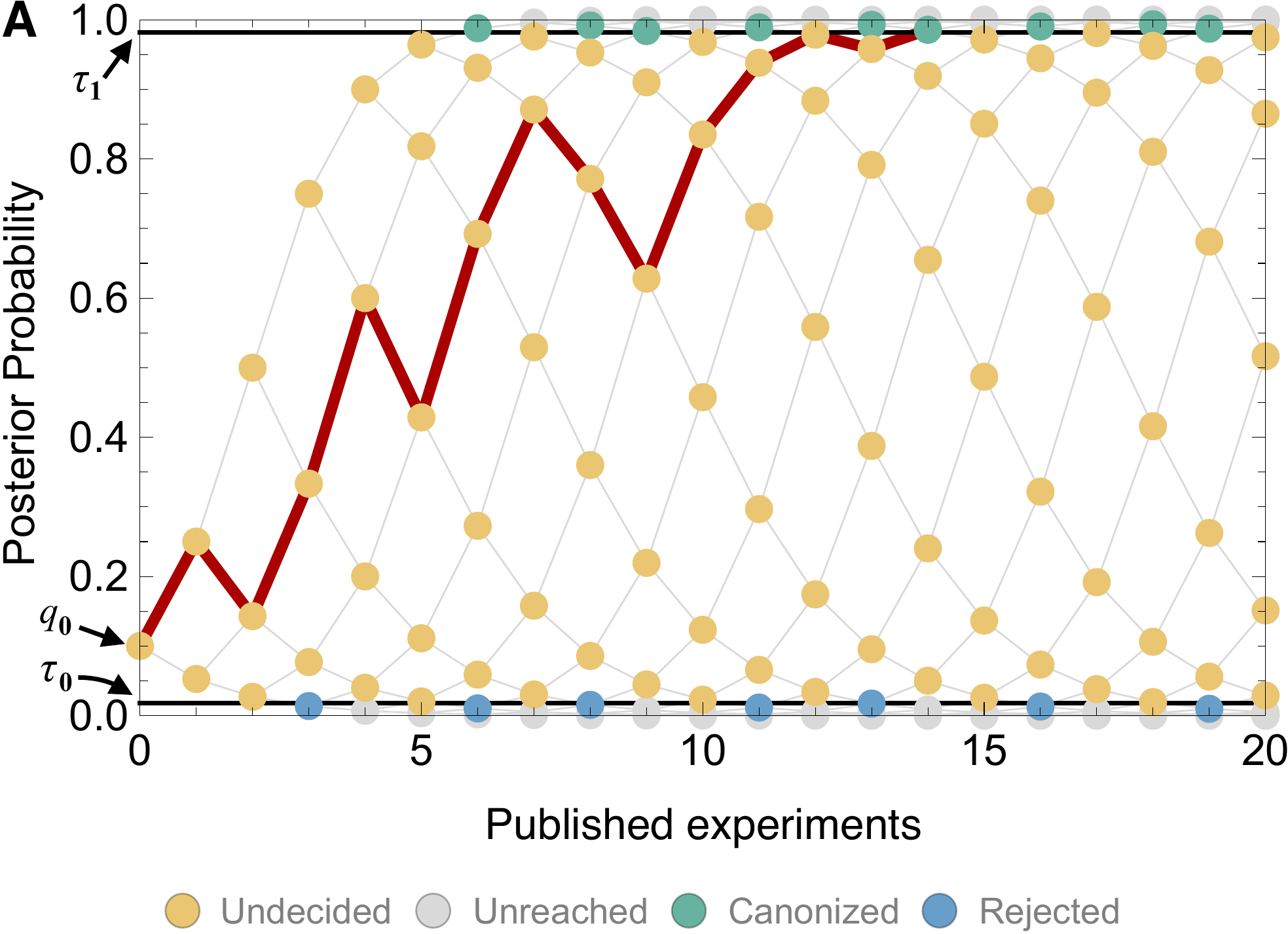}
\hspace{.3in}
\includegraphics[width=3in,valign=t]{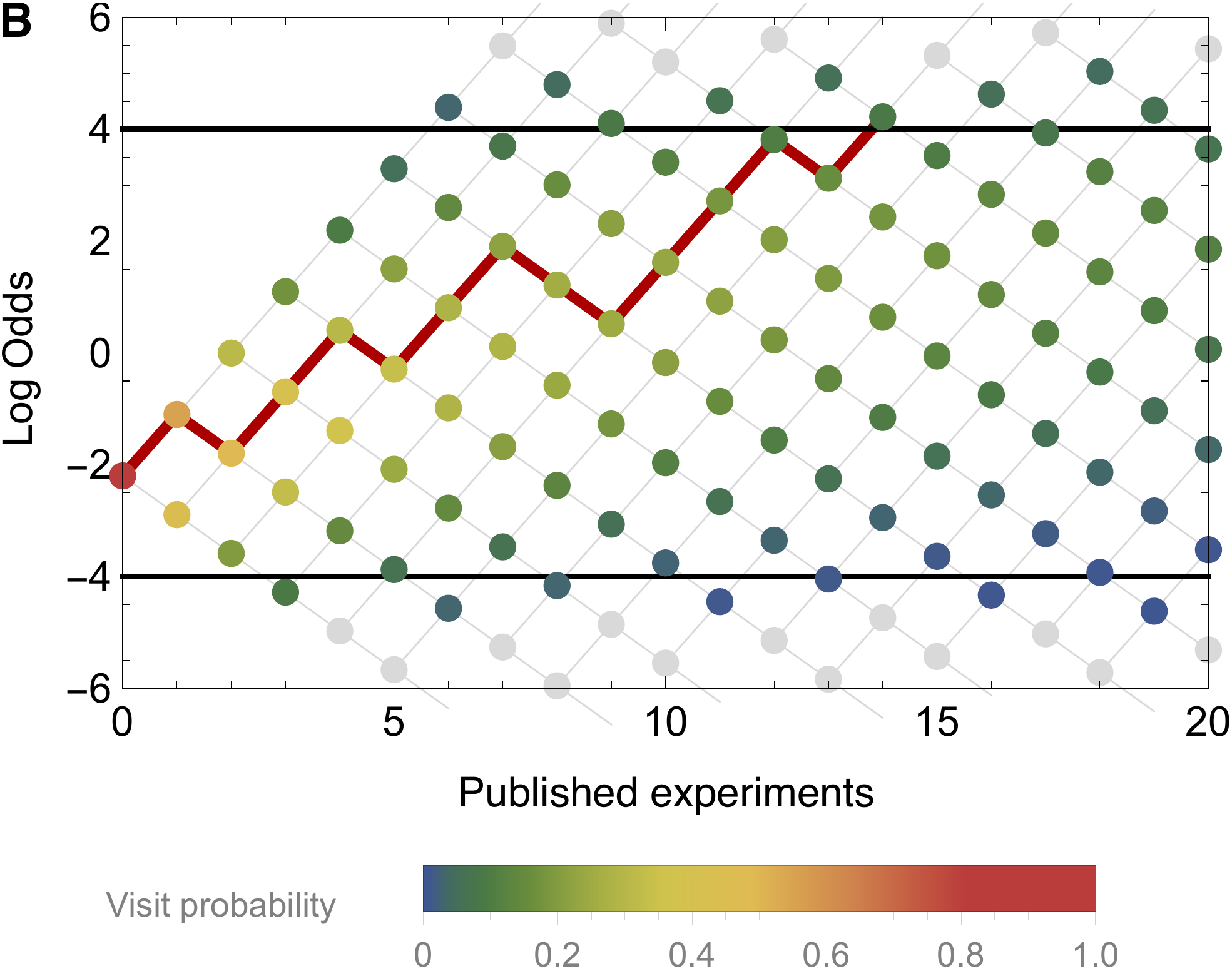}
\caption{{\bf A time-directed graph represents the evolution of belief over time}. In panel A, the horizontal axis indicates the number of experiments published and the vertical axis reflects the observer's belief, quantified as the probability that the claim is true. The process begins at the single point at far left with an initial belief $q_0$. Each subsequent experiment either supports the claim, moving to the next node up and right, or contradicts the claim, moving to the next node down and right. At yellow nodes, the status of the claim is as yet undecided. At green nodes, it is canonized as fact, and at blue nodes, it is rejected as false. The black horizontal lines show the evidentiary standards ($\tau_0$ and $\tau_1$). The red path shows one possible trajectory, in which a positive experiment is followed by a negative, then two positives, then a negative, etc., ultimately becoming canonized as fact when it reaches the upper boundary. Panel B shows the same network, but with the vertical axis representing log odds and using color to indicate the probability that the process visits that node. In log-odds space, each published positive result shifts belief by the constant distance $d_1>0$ and each negative result by a different distance $d_0<0$. Shown here (in both panel A and B) is a false claim with false positive rate $\alpha=0.2$, false negative rate $\beta=0.4$, publication probabilities $p_0=0.1$ and $p_1=1$, and initial belief $q_0=0.1$. In this case, the claim is likely to be canonized as fact, despite being false.}
\label{fig:MarkovChainSchematic}
\end{figure*}

Thus, mathematically, we model belief in the truth of a claim as a discrete-time Markov chain $\left\{q_k\right\}_{k=0}^\infty$ with absorbing boundaries at the evidentiary standards for canonization or rejection (Fig.~\ref{fig:MarkovChainSchematic}A). When the Markov chain represents belief, its possible values lie in the interval from 0 to 1. For mathematical convenience, however, it is often helpful to convert belief to the log odds scale, that is, $\ln(q_k/(1-q_k))$. Some algebra shows that the log odds of belief $q_k(y)$ can be written as
\begin{eqnarray}
\ln\left(\dfrac{q_k(y)}{1-q_k(y)}\right) &=& y \ln \left(\dfrac{1-\beta}{\alpha}\right) + (k-y) \ln \left(\dfrac{\beta}{1-\alpha}\right) \nonumber \\ &&+ \ln \left(\dfrac{q_0}{1-q_0}\right).
\label{eq:log_odds_posterior_belief}
\end{eqnarray}
The log odds scale is convenient because, as eq.~\ref{eq:log_odds_posterior_belief} shows, each published positive outcome increases the log odds of belief by a constant increment
$$
d_1 = \ln\left(\dfrac{1-\beta}{\alpha}\right) > 0
$$ 
(Fig.~\ref{fig:MarkovChainSchematic}B). Each published negative outcome decreases the log odds of belief by 
$$
d_0 = \ln \left(\dfrac{\beta}{1-\alpha}\right) < 0.
$$
Below, we will see that much of the behavior of our model can be understood in terms of the expected change in the log odds of belief for each published outcome. For a true claim, the expected change in the log odds of belief is
\begin{equation}
d_1 \omega_T + d_0 (1-\omega_T)
\label{eq:exp_change_true}
\end{equation}
whereas for a false claim, the expected change in the log odds of belief is 
\begin{equation}
d_1 \omega_F + d_0 (1-\omega_F).
\label{eq:exp_change_false}
\end{equation}

\subsection{Computing canonization and rejection probabilities}

In general, we cannot obtain a closed-form expression for the probability that a claim is canonized as fact or for the probability that it is rejected as false. We can, however, derive recursive expressions for the probabilities that after $k$ published experiments a claim has been canonized as fact, has been discarded as false, or remains undecided. From these, it is straightforward to compute the canonization and rejection probabilities numerically to any desired level of precision. 

For each number of published experiments $k$, the state space for $Y_k$ is simply $Y_k \in \left\{0,1,\ldots,k\right\}$. Partition this state space as follows:
\begin{eqnarray*}
\mathcal{C}_k & = & \left\{y:q_k(y) > \tau_1\right\} \\
\mathcal{I}_k & = & \left\{y:q_k(y) \in \left[\tau_0, \tau_1\right]\right\} \\
\mathcal{R}_k & = & \left\{y:q_k(y) < \tau_0\right\}. 
\end{eqnarray*}
That is, $\mathcal{C}_k$ is the set of outcomes corresponding to a belief greater than the evidentiary standard for canonization, $\mathcal{R}_k$ is the set of outcomes corresponding to a belief less than the evidentiary standard for rejection, and $\mathcal{I}_k$ is the set of outcomes corresponding to belief in between these two standards (the ``interior'').  Let $T$ be the number of publications until a claim is either canonized or rejected. Formally, 
\begin{displaymath}
T = \min\left\{k:Y_k \in \mathcal{C}_k \cup \mathcal{R}_k \right\}.
\end{displaymath}

\begin{figure*}
\centering
\includegraphics[width=.5\linewidth]{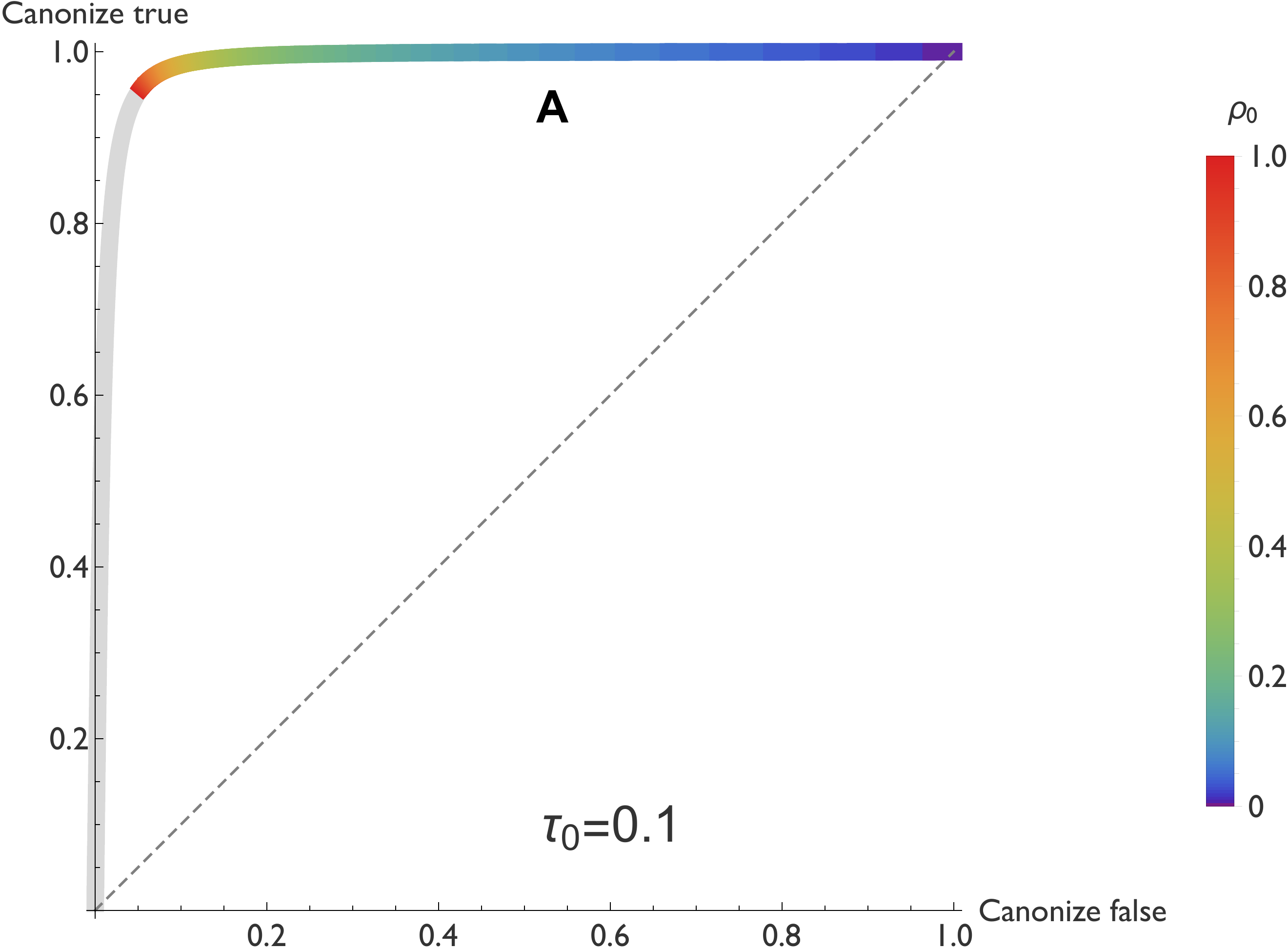}
\includegraphics[width=.45\linewidth]{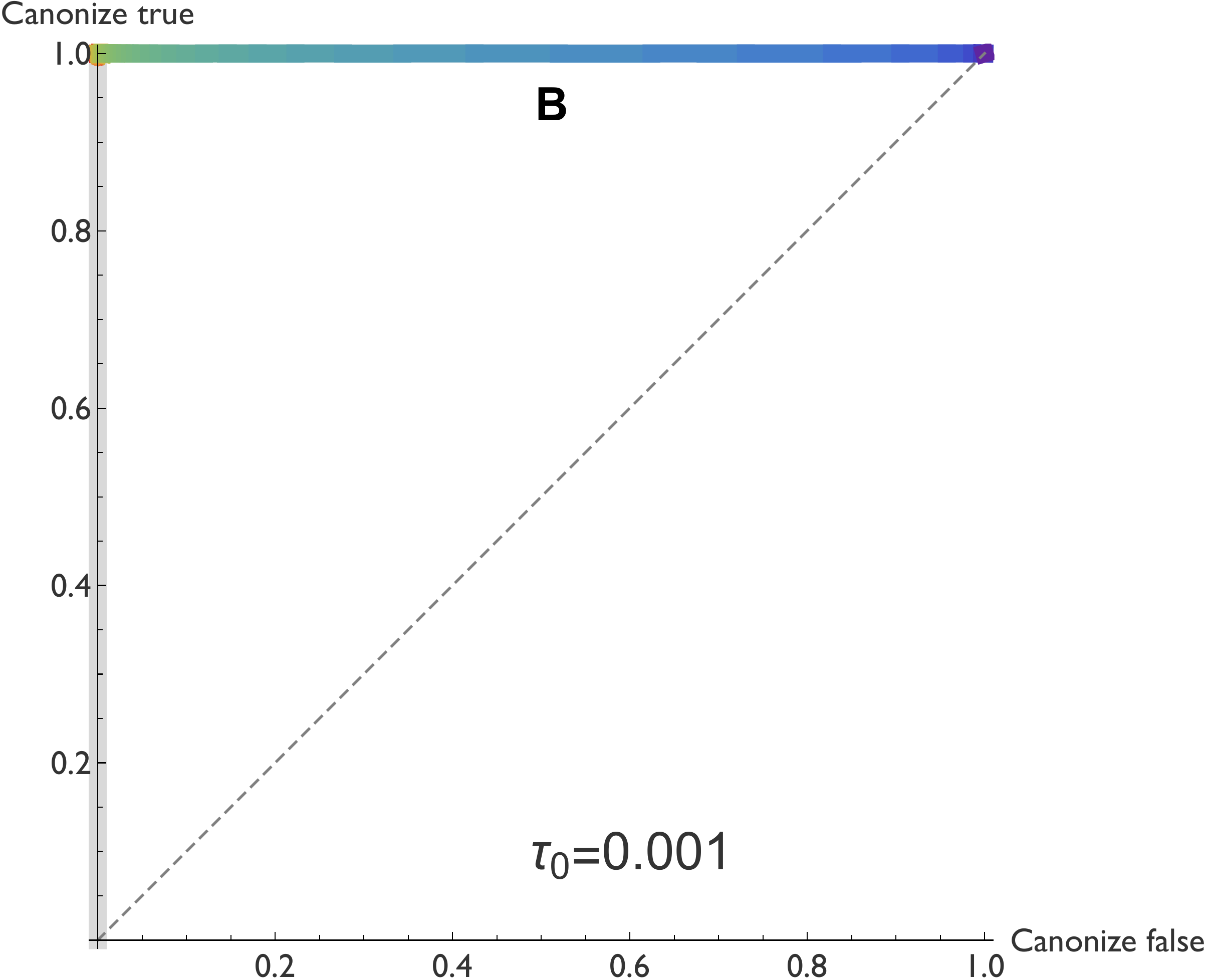}
\caption{{\bf ROC curves reveal that true claims are almost always canonized as fact}. In the receiver operating characteristic (ROC) curves shown here, the vertical axis represents the probability that a true claim is correctly canonized as fact, and the horizontal axis represents the probability that a false one is incorrectly canonized as fact. Panel A: lax evidentiary standards $\tau_0=0.1$ and $\tau_1=0.9$. Panel B: strict evidentiary standards $\tau_0=0.001$ and $\tau_1=0.999$. Error rates and initial belief are $\alpha=0.05$, $\beta=0.2$, and $q_0=0.5$. Each point along the ROC curve corresponds to a different value of negative publication rate, $\rho_0$, as indicated by color. Grey regions of the curve correspond to the unlikely situations in which $\rho_0>\rho_1=1$, i.e., negative results are more likely to be published than positive ones. The figures reveal two important points. First, when negative results are published at any rate $\rho_0\leq 1$, the vast majority of true claims are canonized as fact. Second, when negative results are published at a low rate ($\rho_0$ less than 0.3 or 0.2 depending on evidentiary standards), many false claims will also be canonized as true.}
\label{fig:roc}
\end{figure*}

For a true claim, we calculate the probability of canonization as follows. (A parallel set of equations gives the probability of canonization for a false claim.) For each $y \in \mathcal{I}_k$, define $p_k(y) = \mathrm{Prob}\left\{Y_k=y,T>k\right\}$. That is, $p_k(y)$ is the probability that there are exactly $y$ positive outcomes in the first $k$ publications, and the claim has yet to be canonized or rejected by publication $k$. Suppose these probabilities are known for each $y \in \mathcal{I}_k$. Then for each $y \in \mathcal{I}_{k+1}$, these probabilities can be found recursively by
\begin{displaymath}
p_{k+1}(y) = \omega_T \, p_k(y-1) + (1-\omega_T) p_k(y).
\end{displaymath}
For computational purposes, in the recursion above we define $p_k(y)=0$ whenever $y \notin \mathcal{I}_k$. The probability that the claim has yet to be canonized or rejected by publication $k$ is simply
\begin{displaymath}
\mathrm{Prob}\left\{T>k\right\} = \sum_{y \in \mathcal{I}_k} p_k(y).
\end{displaymath}
Let $\phi_k$ be the probability that a claim is first canonized at publication $k$. Formally,
\begin{displaymath}
\phi_k = \sum_{y: y-1 \in \mathcal{I}_{k-1}\mbox{ and }y \in \mathcal{C}_k} \omega_T \, p_{k-1}(y-1).
\end{displaymath}
Let $k^\star$ be the smallest value of $k$ for which $\mathrm{Prob}\left\{T>k\right\} \leq \epsilon$. To calculate the probability of canonization, we calculate $p_k(y)$ for all $k=1,\ldots,k^\star$. The probability of canonization is then $\sum_{k=1}^{k^\star} \phi_k$. For the analyses in this paper, we have set $\epsilon=10^{-4}$.

\section{Results}

We focus throughout the paper on the dynamic processes by which false claims are canonized as facts, and explore how the probability of this happening depends on properties of the system such as the publication rate of negative results, the initial beliefs of researchers, the rates of experimental error, and the degree of evidence required to canonize a claim. In principle, the converse could be a problem as well: true claims could be discarded as false. However, this is rare in our model. Publication bias favors the publication of positive results and therefore will not tend to cause true claims to be discarded as false, irrespective of other parameters. We first establish this, and then proceed to a detailed examination of how scientific experimentation and publication influences the rate at which false claims are canonized as fact. 

\subsection{True claims tend to be canonized as facts}

In our model, true claims are almost always canonized as facts. Figure \ref{fig:roc} illustrates this result in the form of a receiver operating characteristic (ROC) curve. Holding the other parameters constant, the curve varies the negative publication rate $\rho_0$, and uses the vertical and horizontal axes to indicate the probabilities that true and false claims respectively are canonized as fact.

One might fear that as the probability $\rho_0$ of publishing negative results climbs toward unity, the risk of rejecting a true claim would increase dramatically as well. This is not the case. Even as the probability of publishing negative results approaches 1, the risk of rejecting a true claim is low when evidentiary standards are lax (Fig.~\ref{fig:roc}A), and negligible when evidentiary standards are strict (Fig.~\ref{fig:roc}B).

This result turns out to be general across a broad range of parameters. Assuming the mild requirements that (i) tests of a true claim are more likely to result in positive publications than negative publications (i.e., $\omega_T > 1/2$, or equivalently $(1-\beta) > \beta \rho_0$), and (ii) positive published outcomes increase belief that the claim is true ($d_1 > 0$, or equivalently $(1-\beta) > \alpha$), true claims are highly likely to be canonized as facts. The exceptions occur only when minimal evidence is needed to discard a claim, i.e., when initial belief is small ($q_0 \approx \tau_0$). In such cases a bit of bad luck---the first one or two published experiments report false negatives, for example---can cause a true claim to be rejected. But otherwise, truth is sufficient for canonization. 

Unfortunately, truth is not required for canonization. The risk of canonizing a false claim---shown on the horizontal axis value in figure \ref{fig:roc}---is highly sensitive to the rate at which negative results are published. When negative results are published with high probability, false claims are seldom canonized. But when negative results are published with lower probability, many false claims are canonized. 

Thus we see that the predominant risk associated with publication bias is the canonization of false claims. In the remainder of this analysis, we focus on this risk of incorrectly establishing a false claim as a fact.

\begin{figure}[!t]
\centering
\includegraphics[width=\columnwidth]{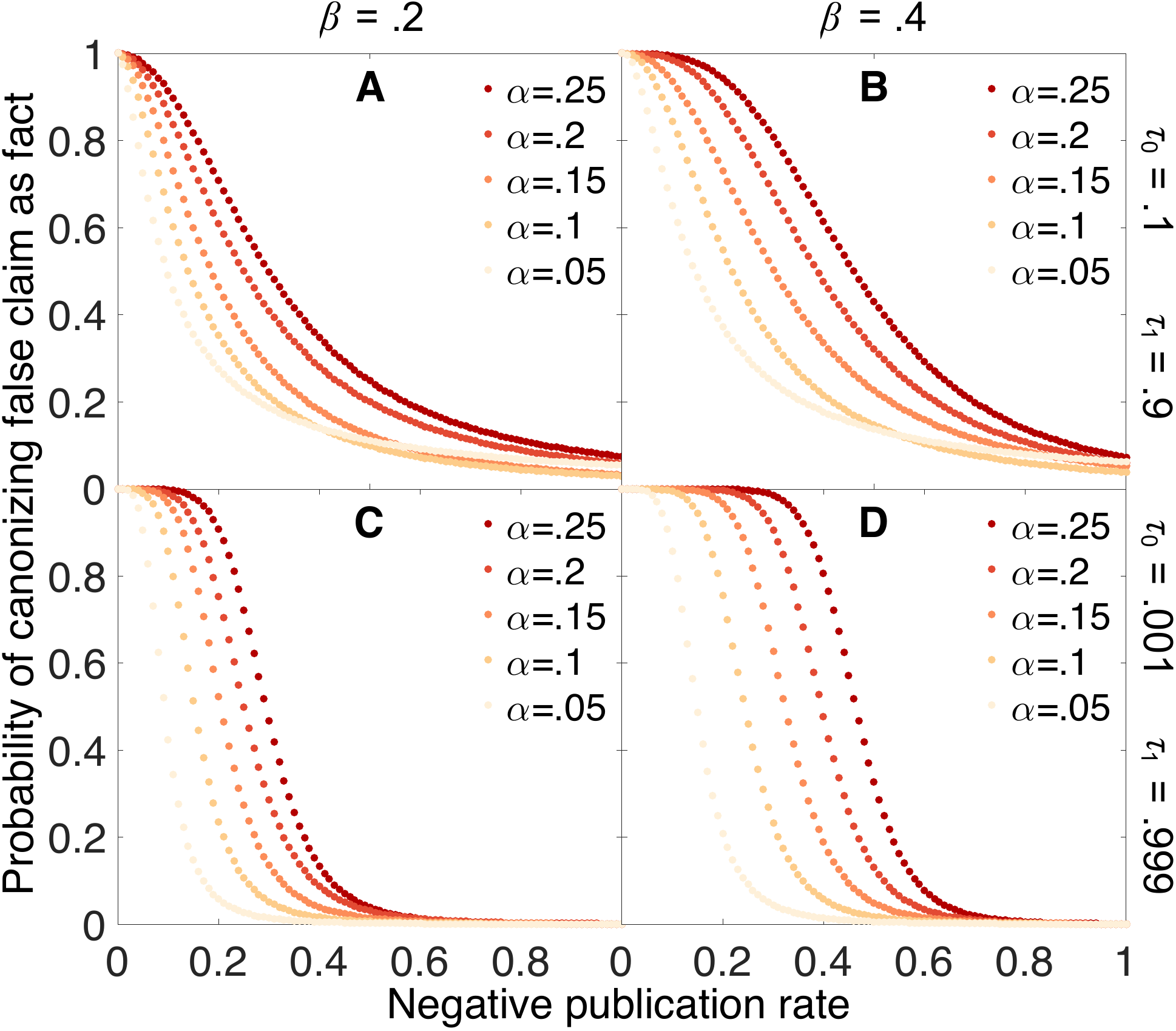}
\caption{{\bf Publishing negative outcomes is essential for rejecting false claims}. Probability that a false claim is incorrectly canonized, as a function of the negative publication rate. Throughout, initial belief is $q_0 = 0.5$, and individual data series show false positive rates $\alpha=0.05$ (yellow), $0.10,\ldots,0.25$ (red). Top row: weak evidentiary standards $\tau_0=0.1$ and $\tau_1=0.9$. Panel A: false negative rate $\beta=0.2$. Panel B: $\beta=0.4$. Panels C--D: similar to panels A--B, with more demanding evidentiary standards $\tau_0=0.001$ and $\tau_1=0.999$.} 
\label{fig:NegPubRate}
\end{figure}

\subsection{Publication of negative results is essential}

As we discussed in the introduction, authors and journals alike tend to be reluctant to publish negative results, and as we found in the previous subsection, when most negative results go unpublished, science performs poorly. Here, we explore this relationship in further detail. 

Figure \ref{fig:NegPubRate} shows how the probability of erroneously canonizing a false claim as fact depends on the probability $\rho_0$ that a negative result is published. False claims are likely to be canonized below a threshold rate of negative publication, and unlikely to be canonized above this threshold. For example, when the false positive rate $\alpha$ is 0.05, the false negative rate $\beta$ is 0.4, and the evidentiary requirements are strong (yellow points in Panel \ref{fig:NegPubRate}D), a false claim is likely to be canonized as fact unless negative results are at least 20\% as likely as positive results to be published. 

Figure~\ref{fig:NegPubRate} also reveals that the probability of canonizing false claims as facts depends strongly on both the false positive rate and the false negative rate of the experimental tests. As these error rates increase, an increasingly large fraction of negative results must be published to preserve the ability to discriminate between true and false claims. 

\begin{figure}[!t]
\centering
\includegraphics[width=\columnwidth]{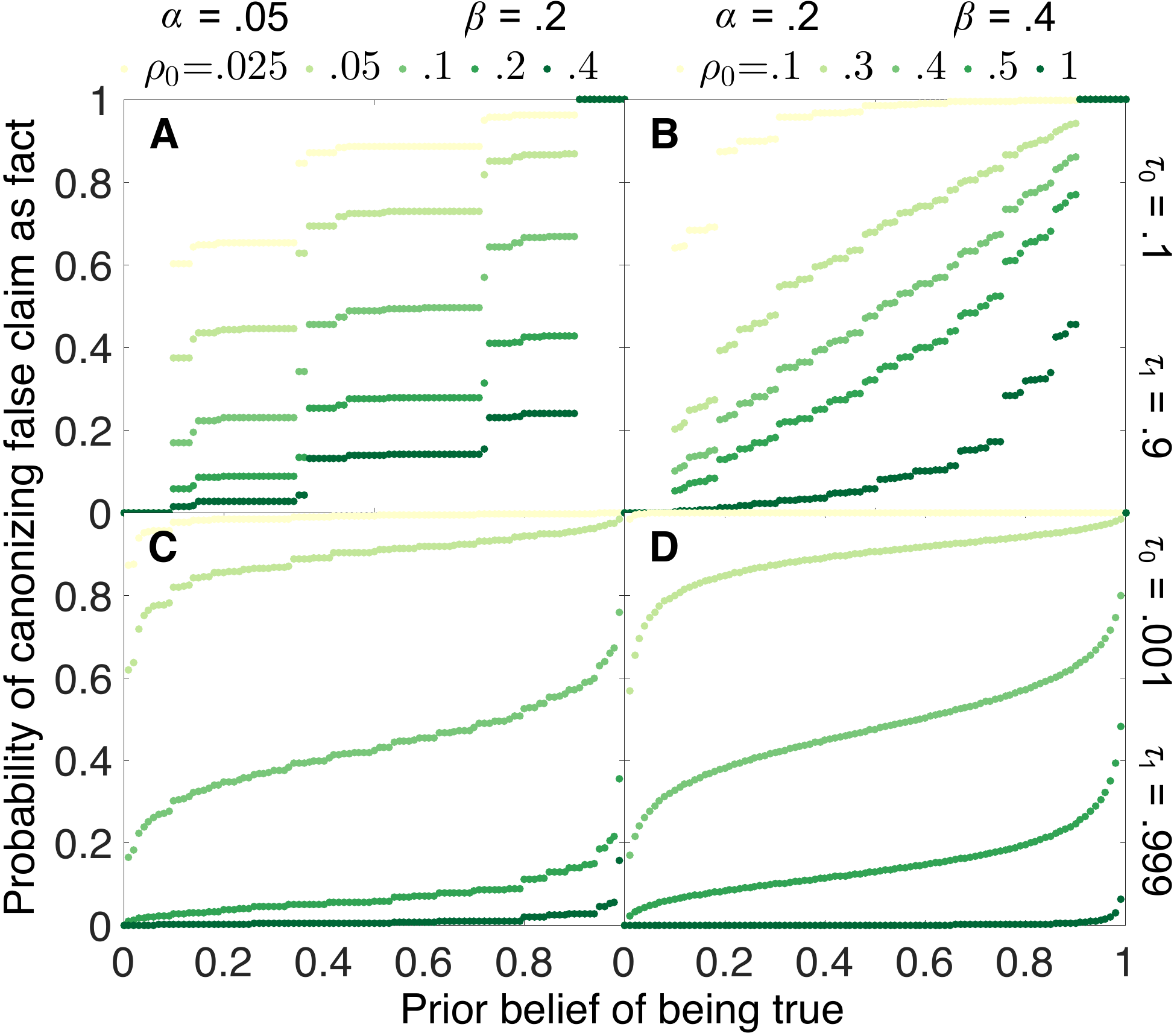}
\caption{{\bf False canonization rates are relatively insensitive to initial belief, unless experimental tests are inaccurate and evidentiary standards are weak}. Probability that a false claim is mistakenly canonized as a true fact vs.~prior belief for various negative publication rates. Top row: weak evidentiary standards $\tau_0=0.1$ and $\tau_1=0.9$. Panel A: false positive rate $\alpha=0.05$, false negative rate $\beta=0.2$, and publication rate of negative results $\rho_0=0.025$ (light green), $0.05,0.1,0.2,0.4$ (dark green). Panel B: $\alpha=0.2$, $\beta=0.4$, and $\rho_0=0.1$ (light green), $0.3,0.4,0.5,1$ (dark green). Panels C--D: similar to panels A--B, with more demanding evidentiary standards $\tau_0=0.001$ and $\tau_1=0.999$.} 
\label{fig:InitialBeliefs}
\end{figure}

\subsection{Initial beliefs usually do not matter much}

If the scientific process is working properly, it should not automatically confirm what we already believe, but rather it should lead us to change our beliefs based on evidence. Our model indicates that in general, this is the case. 

Figure \ref{fig:InitialBeliefs} shows how the probability that a false claim is canonized as true depends on the initial belief $q_0$ that the claim is true. Under most circumstances, the probability of canonization is relatively insensitive to initial belief. False canonization rates depend strongly on initial belief only when evidentiary standards are weak and experiments are highly prone to error (Fig.~\ref{fig:InitialBeliefs}B). In this case, belief is a random walk without a systematic tendency to increase or decrease with each published outcome, and thus the odds of canonization or rejection depend most strongly on the initial belief. 

The step-function-like appearance of some of the results in Fig.~\ref{fig:InitialBeliefs}, particularly Fig.~\ref{fig:InitialBeliefs}A, is a real property of the curves in question and not a numerical artifact. The ``steps'' arise because, when evidentiary standards are weak, canonization or rejection often happens after a small number of experiments. Because the number of experiments must be integral, probabilities of false canonization can change abruptly when a small change in initial belief increases or decreases the number of experiments in the most likely path to canonization or rejection.

\begin{figure}[!t]
\centering
\includegraphics[width=\columnwidth]{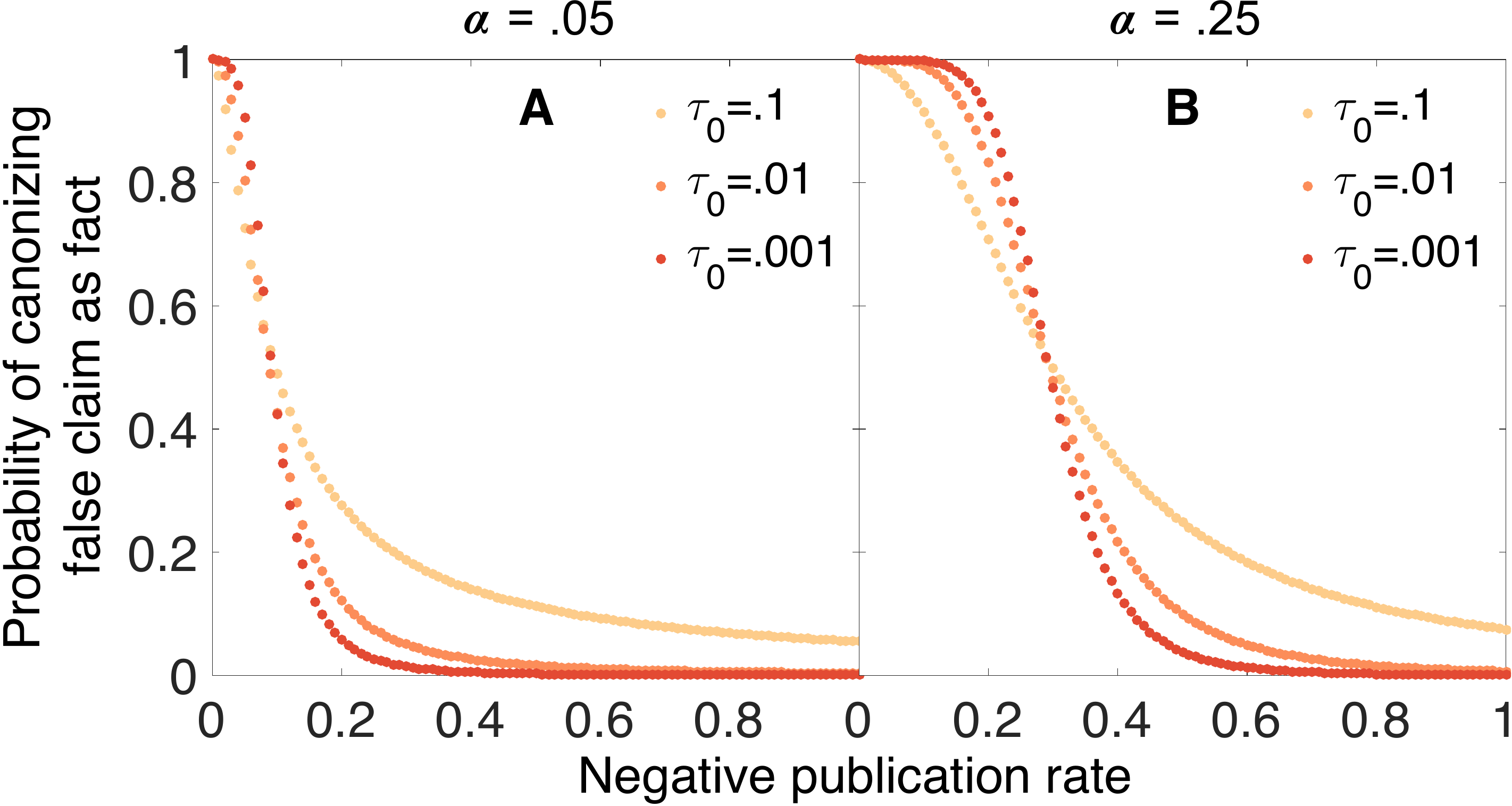}
\caption{{\bf Strengthening evidentiary requirements does not necessarily decrease canonization of false facts.} In panel \ref{fig:EvidentiaryStandards}A, the false positive rate is $\alpha=0.05$, the false negative rate is $\beta=0.2$, the original belief in the claim is $q_0=0.5$, and the evidentiary standards are symmetric $\tau_1 = 1 - \tau_0$. In panel \ref{fig:EvidentiaryStandards}B, the false positive rate is increased to $\alpha=0.25$ while the other parameters remain unchanged. Particularly in this latter case, increasing evidentiary standards does not necessarily decrease the rate at which false claims are canonized as facts.}
\label{fig:EvidentiaryStandards}
\end{figure}

\subsection{Stronger evidentiary standards do not reduce the need to publish negative outcomes}

We have seen in the previous sections that the scientific process struggles to distinguish true from false claims when the rate of publishing negative results is low. We might hope that we could remedy this problem simply by demanding more evidence before accepting a claim as fact. Unfortunately, this is not only expensive in terms of time and effort---sometimes it will not even help.

Figure \ref{fig:EvidentiaryStandards} illustrates the problem. In this figure, we see the probability of canonizing a false claim as a function of negative publication rate for three different evidentiary standards: $\tau_0=0.1$, $\tau_0=0.01$, and $\tau_0=0.001$. When the false positive rate $\alpha$ is relatively low (Fig.~\ref{fig:EvidentiaryStandards}A), increasing the evidentiary requirements reduces the chance of canonizing a false claim for negative publication rates above 0.1 or so, but below this threshold there is no advantage to requiring stronger evidence. When the false positive rate is higher (Fig.~\ref{fig:EvidentiaryStandards}B), the situation is even worse: for negative publication rates below 0.3 or so, increasing evidentiary requirements actually increases the chance of canonizing a false fact. 

\begin{figure}[!t]
\centering
\includegraphics[width=\columnwidth]{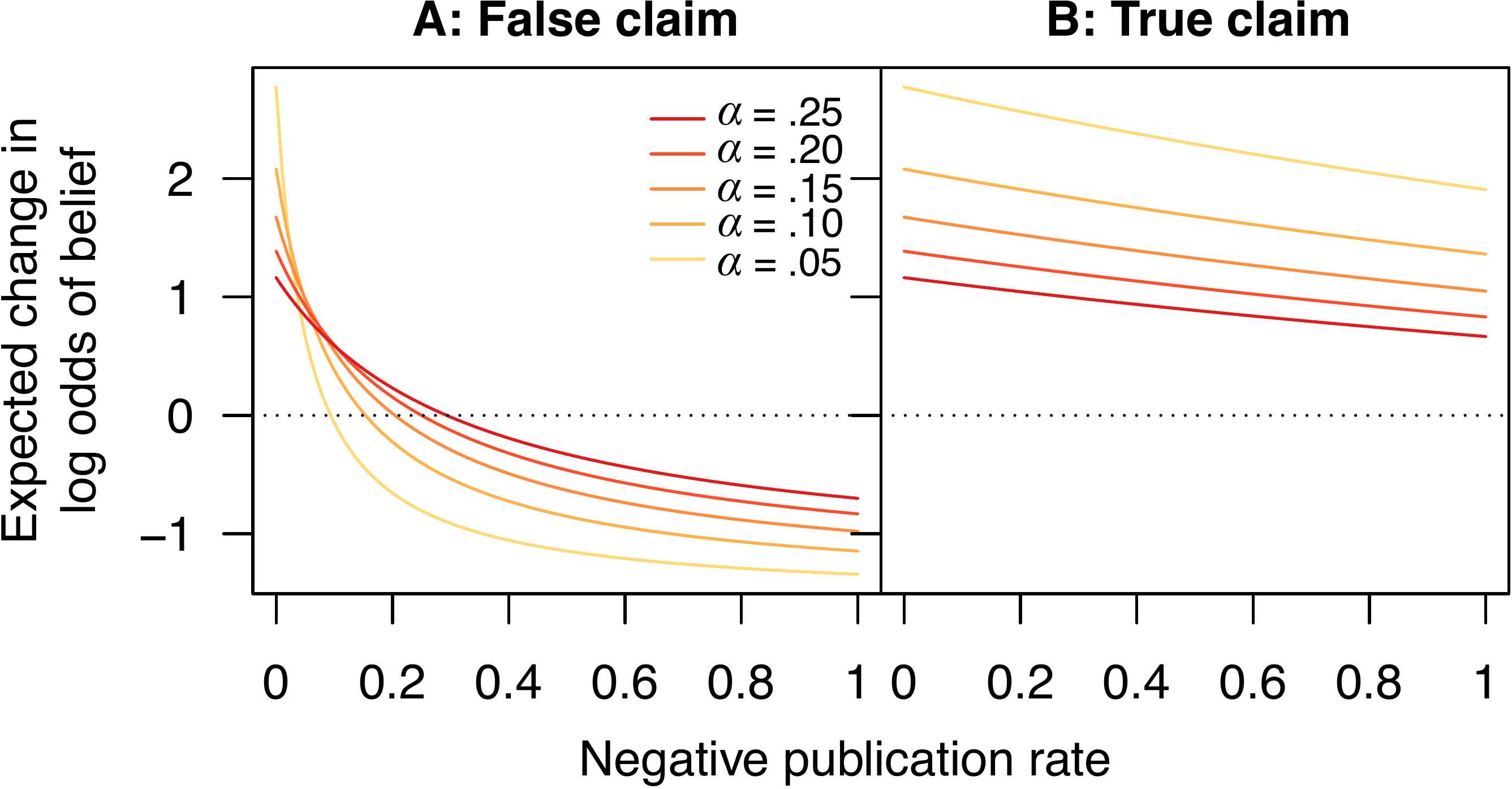}
\caption{{\bf Scientific activity will tend to increase belief in false claims if too few negative outcomes are published}. Expected change in log odds of belief vs.~negative publication rate for (A) false and (B) true claims. Lines show false positive rates $\alpha=0.05$ (yellow), $0.10,\ldots,0.25$ (red). Other parameter values are false negative rate $\beta=0.2$ and positive publication rate $\rho_1=1$.}
\label{fig:ExpectedChangeInLogOdds}
\end{figure}

The limited benefits of strengthening evidentiary standards can be understood through the mathematical theory of random walks \cite{norris1998markov}. In short, the thresholds of belief for canonizing or rejecting a claim are {\em absorbing boundaries} such that once belief attains either boundary, the walk terminates and beliefs will not change further. Increasing the evidentiary standards for canonization or rejection is tantamount to increasing the distance between these boundaries and the initial belief state. Basic results from the theory of random walks suggest that, as the distance between the initial state and the boundaries increases, the probability of encountering one boundary before the other depends increasingly strongly on the average change in the log odds of belief at each step (experiment), and less on the random fluctuations in belief that arise from the stochasticity of the walk. Thus, for exacting evidentiary standards, the probability of eventual canonization or rejection depends critically on the average change in the log odds of belief for each experiment. These are given by eq.~\ref{eq:exp_change_true} for a true claim and eq.~\ref{eq:exp_change_false} for a false one.
	
Figure~\ref{fig:ExpectedChangeInLogOdds} shows how the expected change in log odds of belief varies in response to changes in the publication rate of negative outcomes, for both false and true claim. Critically, for false claims, if too few negative outcomes are published, then on average each new publication will \emph{increase} the belief that the claim is true, because there is a high probability this publication will report a positive result. Thus, paradoxically, scientific activity does not help sort true claims from false ones in this case, but instead promotes the erroneous canonization of false claims. The only remedy for this state of affairs is to publish enough negative outcomes that, on average, each published result moves belief in the ``correct'' direction, that is, towards canonization of true claims (a positive average change per experiment in log odds of belief) and rejection of false ones (a negative average change per experiment in log odds of belief).

Two additional points are in order here. First, for true claims, under most circumstances the expected change in the log odds of belief is positive (Fig.~\ref{fig:ExpectedChangeInLogOdds}B). That is, on average, scientific activity properly increases belief in true claims, and thus the risk of incorrectly rejecting a true claim is small (under reasonable evidentiary standards). Second, the observation that more exacting evidentiary standards can occasionally increase the chance of incorrectly canonizing a false claim is not much of an argument in favor of weaker evidentiary standards. In short, weaker standards cause canonization or rejection to depend more strongly on the happenstance of the first several published experiments. When scientific activity tends to increase belief in a false claim, weaker evidentiary standards appear beneficial because they increase the chance that a few initial published negatives will lead to rejection and bring a halt to further investigation. While this is a logical result of the model, it is somewhat tantamount to stating that, if scientific activity tends to increase belief in false claims, then the best option is to weaken the dependence on scientific evidence. More robust practices for rejecting false claims seem desirable.

\subsection{$P$-hacking dramatically increases the probability of canonizing false claims}

Our model has been based on the optimistic premise that the significance levels reported in each study accurately reflect the actual false positive rates. This means that there is only a 5\% chance that a false claim will yield a positive result at the $\alpha=0.05$ level.

\begin{figure}
\centering
\includegraphics[width=.95\columnwidth]{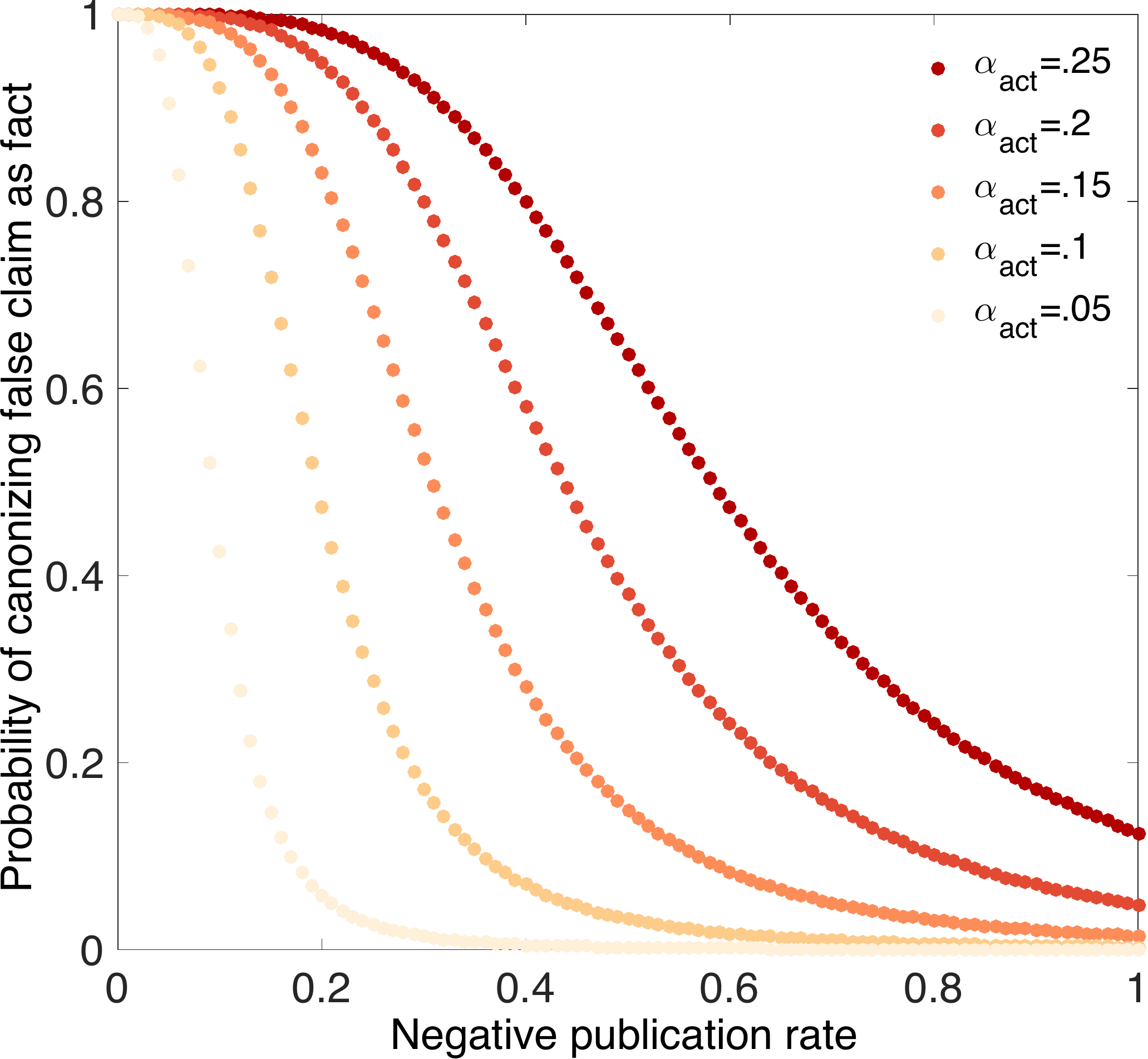}
\caption{{\bf $p$-hacking dramatically increases the chances of canonizing false claims}. Probability that a false claim is canonized as fact vs.~fraction of negative outcomes. Throughout, all positive outcomes are published ($p_1=1$), and the nominal false positive rate is $\anom=0.05$, the false negative rate is $\beta=0.2$, and evidentiary standards are strong ($\tau_0 = 0.001$ and $\tau_1 = 0.999$). Curves show actual false positive rates $\aact=0.05$ (yellow), $0.10,\ldots,0.25$ (red). Compared with Fig.~\ref{fig:NegPubRate}C, in which the nominal rates are equal to the actual rates, the probability of canonizing a false claim as fact is substantially higher.}
\label{fig:pHacking}
\end{figure}

In practice, reported significance levels can be misleading. Questionable research practices of all sorts can result in higher-than-reported false positive rates; these include p-hacking \citep{head2015extent}, outcome switching \citep{le2015restoring}, unreported multiple comparisons \citep{tannock1996false}, data dredging \citep{smith2002data}, HARKing---hypothesizing after the results are known \citep{kerr1998harking}, data-dependent analysis \citep{gelman2014statistical}, and opportunistic stopping or continuation \citep{pocock1987statistical}. Insufficient validation of new technologies, or even software problems can also drive realized error rates far above what is expected given stated levels of statistical confidence (see e.g.~ref.~\cite{eklund2016cluster}).
Research groups may be positively disposed toward their prior hypotheses or reluctant to contradict the work of closely allied labs. Finally, industry-sponsored clinical trials often allow the sponsors some degree of control over whether results are published \cite{kasenda2016agreements}, resulting in an additional source of publication bias separate from the journal acceptance process. 

To understand the consequences of these problems and practices, we can extend our model to distinguish the actual false positive rate \aact~from nominal false positive rate \anom~which is reported in the paper and used by readers to draw their inferences. We assume the actual false positive rate is always at least as large as the nominal rate, that is, $\aact \geq \anom$. In this scenario, the probability that a false claims leads to a positive published outcome depends on the actual false positive rate, i.e., 
\[\omega_F=\frac{\aact \rho_1}{\aact \rho_1 + (1-\aact)\rho_0}.\] 
However, the change in belief following a positive or negative published outcome respectively depends on the nominal false positive rate:
\begin{eqnarray}
d_1 &=& \ln\left(\frac{1-\beta}{\anom}\right) \nonumber \\
d_0 &=& \ln \left(\frac{\beta}{1-\anom}\right)\nonumber
\end{eqnarray}

An inflated false positive rate makes it much more likely that false claims will be canonized as true facts (Fig.~\ref{fig:pHacking}). For example, suppose the false negative rate is $\beta=0.2$, the nominal false positive rate is $\anom=0.05$, but the actual false positive rate is $\aact=0.25$. Even eliminating publication bias against negative outcomes (i.e., $\rho_0=1$) and using strong evidentiary standards does not eliminate the possibility that false claims will be canonized as facts under these circumstances (Fig.~\ref{fig:pHacking}). Less dramatic inflation of the false positive rate leaves open the possibility that true vs.~false claims can be distinguished, but only if a higher percentage of negative outcomes are published. 

\subsection{Increasing negative publication rates as a claim approaches canonization greatly increases accuracy}

Thus far we have told a troubling story. Without high probabilities of publication for negative results, the scientific process may perform poorly at distinguishing true claims from false ones. And there are plenty of reasons to suspect that negative results may not always be likely to be published. 
 
However, authors, reviewers, and editors are all drawn to unexpected results that challenge or modify prevalent views---and for a claim widely believed to be true, a negative result from a well-designed study is surprising. As a consequence, the probability of publishing a negative result may be higher for claims that are already considered likely to be true \cite{silvertown1997does,ioannidis2005early}

In a simulation of point estimation by successive experimentation, de Winter and Happee considered an even more extreme situation in which it is only possible to publish results that contradict the prevailing wisdom \cite{de2013selective}. They argue that this has efficiency benefits, but their results have been challenged persuasively by van Assen and colleagues \cite{van2014publishing}. In any event, such a publication strategy would not work in the framework we consider here, because a claim could neither be canonized nor rejected if each new published result were required to contradict the current beliefs of the community.

Some meta-analyses have revealed patterns consistent with this model \cite{poulin2000manipulation}. For example, when the fluctuating asymmetry hypothesis was proposed in evolutionary ecology, the initial publications exclusively reported strong associations between symmetry and attractiveness or mating success. As time passed, however, an increasing fraction of the papers on this hypothesis reported negative findings with no association between these variables \cite{simmons1999fluctuating}. A likely interpretation is that initially journals were reluctant to publish results inconsistent with the hypothesis, but as it became better established, negative results came to be viewed as interesting and worthy of publication \cite{simmons1999fluctuating,palmer2000quasireplication,jennions2002relationships}.

\begin{figure}
\centering
\includegraphics[width=\columnwidth]{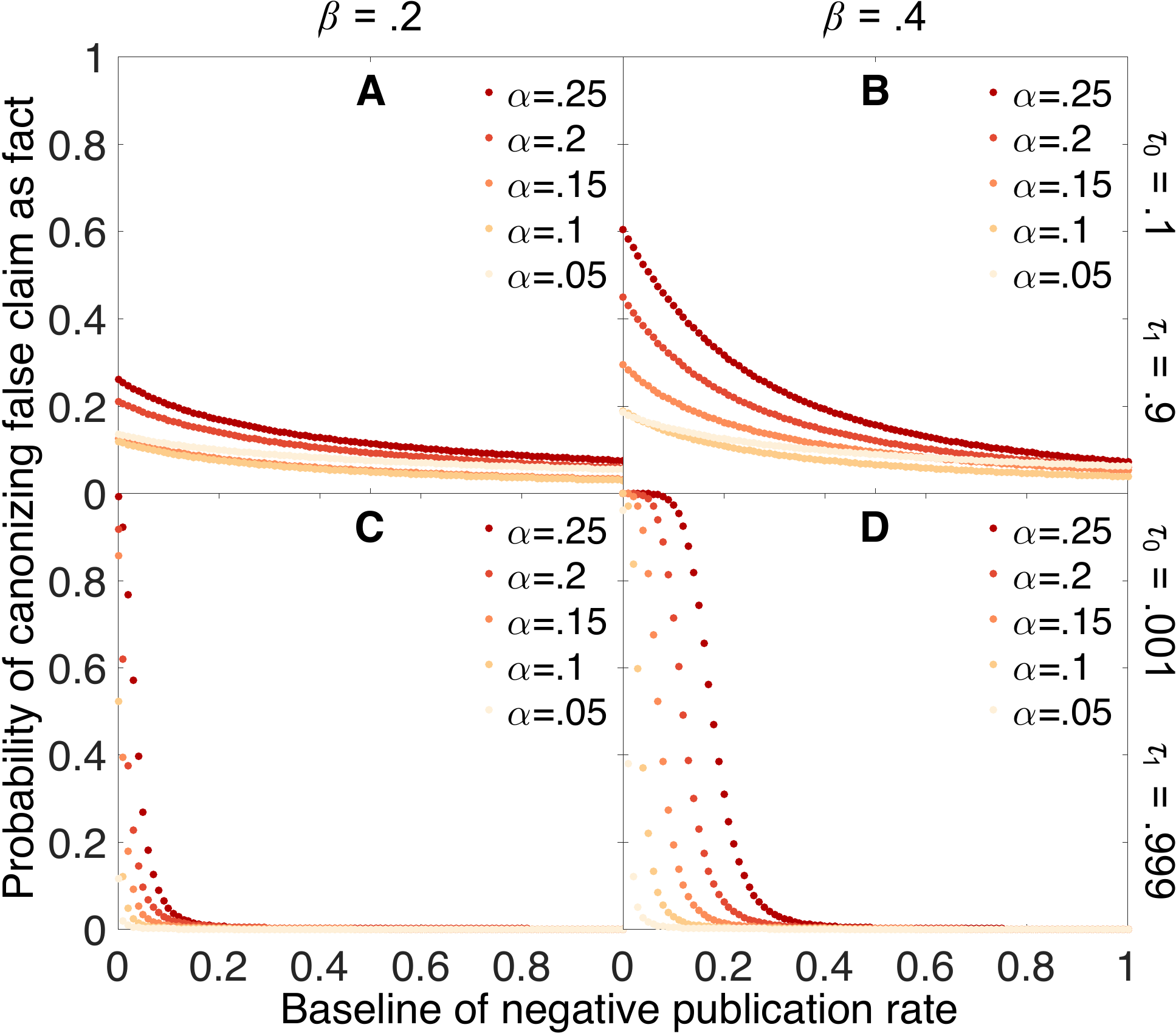}
\caption{{\bf Publishing a larger fraction of negative outcomes as belief increases lessens the chances of canonizing false claims}. Probability that a false claim is mistakenly canonized as a true fact vs.\ baseline probability of publishing a negative outcome. The baseline probability of publishing a negative outcome is the probability that prevails when belief in the claim is weak. The actual probability of publishing a negative outcome increases linearly from the baseline rate when belief is 0 to a value of 1 when belief is 1. All other parameters are the same as in Fig.~\ref{fig:NegPubRate}.}
\label{fig:DynamicRho0}
\end{figure}

To explore the consequences of this effect, we consider a model in which the probability of publishing a negative outcome increases linearly from a baseline value $\rho_b$ when belief in the claim is weak, to a maximum value of $\rho_0=1$ when belief in the claim is strong. We assume that the probability of publishing a negative outcome is $\rho_0 = \rho_b + q(1 - \rho_b)$, where $\rho_b$ is the baseline probability for publishing negative outcomes, and $q$ is the current belief. As before, our agents are unaware of any publication bias in updating their own beliefs.

Figure \ref{fig:DynamicRho0} indicates that dynamic publication rates can markedly reduce (though not eliminate) the false canonization rate under many scenarios. In particular, Fig.~\ref{fig:DynamicRho0} suggests that even if it is difficult to publish negative outcomes for claims already suspected to be false, we can still accurately sort true claims from false ones provided that negative outcomes are more readily published for claims nearing canonization. In practice, this mechanism may play an important role in preventing false results from becoming canonized more frequently.

\section{Discussion}

In the model of scientific inquiry that we have developed here, publication bias creates serious problems. While true claims will seldom be rejected, publication bias has the potential to cause many false claims to be mistakenly canonized as facts. This can be avoided only if a substantial fraction of negative results are published. But at present, publication bias appears to be strong, given that only a small fraction of the published scientific literature presents negative results. Presumably many negative results are going unreported. While this problem has been noted before \cite{knight2003negative}, we do not know of any previous formal analysis of its consequences regarding the establishment of scientific facts. 

\subsection{Should scientists publish all of their results?}

There is an active debate over whether science functions most effectively when researchers publish all of their results, or when they publish only a select subset of their findings \cite{nelson2012let,de2013selective,van2014publishing,mcelreath2015replication}. In our model, we observe no advantage to selective publication; in all cases treated we find that false canonization decreases monotonically with increasing publication of negative results. This seems logical enough. Decision theory affirms that in the absence of information costs, withholding information cannot on average improve performance in a decision problem such as the classification task we treat here \cite{savage1954foundations,good1967principle,ramsey1990weight}. As Good \cite{good1967principle} notes, a decision-maker ``should use all the evidence {\em already} \/available, provided that the cost of doing so is negligible.'' 

Nonetheless, several research groups have argued that selective publishing can be more efficient than publishing the results of all studies. Clearly they must be implicitly or explicitly imposing costs of some sort on the acts of publishing papers or reading them, and it can be instructive to see where these costs lie. One source of such costs comes simply from the increased volume of scientific literature that ensues when all results are published; this is sometimes known as the ``cluttered office'' effect \cite{nelson2012let}. If we envision that search costs increase with the volume of literature, for example, it may be beneficial not to publish everything. 

Another possible cost is that of actually writing a paper and going through the publication process. If preparing a paper for publication is costly relative to doing the experiments which would be reported, it may be advantageous to publish only a subset of all experimental results. This is the argument that de Winter and Happee make when, in a mathematical model of point estimation, they find that selective publication minimizes the variance {\em given the number of publications} (as opposed to the number of experiments conducted). Note, however, that they assume a model of science in which experiments are only published when they contradict the prevailing wisdom---and that their results have been roundly challenged in a followup analysis \cite{van2014publishing}.

McElreath and Smaldino \cite{mcelreath2015replication} analyzed a model that is more similar to ours in structure. As we do, they consider repeated tests of binary-valued hypotheses. But rather than focusing on a single claim at a time, they model the progress of a group of scientists testing a suite of hypotheses. Based on this model, McElreath and Smaldino conclude that there can be advantages to selective publication under certain conditions.

While selective publication certainly can ameliorate the cluttered office problem---observed in their model as the flocking of researchers to questions already shown likely to be false---we are skeptical about the other advantages to selective publication. McElreath and Smaldino's model and results appear to rely in part on their assumption that ``the only information relevant for judging the truth of a hypothesis is its {\em tally}, the difference between the number of published positive and the number of positive negative findings'' (p.~3).

As a mathematical claim, this is incorrect. Presumably the claim is instead intended to be a tactical assumption that serves to simplify the analysis. But this assumption is severely limiting. The tally is often an inadequate summary of the evidence in favor of a hypothesis. One can see the problem with looking only at the tally by considering a simple example in which false positive rates are very low, false negative rates are high, and all studies are published. There is mild evidence that a hypothesis is false if no positive studies and one negative study have been published, but there is strong evidence that the hypothesis is true if three positive and four negative studies have been published. Yet both situations share the same tally: $-1$. The same problem arises when publication bias causes positive and negative findings to be published at different rates.

If one is forced to use only the tally to make decisions, an agent can sometimes make better inferences by throwing away some of the data (i.e., under selective publication). For example, when false negatives are common it may be beneficial to suppress some fraction of the negative results lest they swamp any positive signal from true positive findings. This is not the case when the agent has access to complete information about the number of positive and the number of negative results published. As a result, it is unclear whether most of McElreath and Smaldino's arguments in favor of selective publication are relevant to optimal scientific inference, or whether they are consequences of the assumption that readers draw their inferences based on the tally alone.

\subsection{What do we do about the problem of publication bias?}

Several studies have indicated that much of the publication bias observed in science can be attributed to authors not writing up null results, rather than journals rejecting null results \cite{dickersin1992factors,olson2002publication, franco2014publication}. This does not necessarily exonerate the journals; authors may be responding to incentives that the journals have put in place \cite{song2000publication}. Authors may be motivated by other reputational factors as well. It would be a very unusual job talk, promotion seminar, or grant application that was based primarily upon negative findings. 

So what can we as a scientific community do? How can we avoid canonizing too many false claims, so that we can be confident in the veracity of most scientific facts? In this paper, we have shown that strengthening evidentiary standards does not necessarily help. In the presence of strong publication bias, false claims become canonized as fact not so much because of a few misleading chance results, but rather because on average, misleading results are more likely to be published than correct ones. 

Fortunately, this problem may be ameliorated by several current aspects of the publication process. In this paper, we have modeled claims that have only one way of generating ``positive'' results. For many scientific claims, e.g.~those like our Dicer example that propose particular mechanisms, this may be appropriate. In other cases, however, results may be continuous: not only do we care whether variables $X$ and $Y$ are correlated, but also we want to know about the strength of the correlation, for example. This does not make the problem go away, if stronger or highly significant correlations are seen as more worthy of publication than weaker or non-significant correlations. However, one advantage of framing experimental results as continuous-valued instead of binary is that there may be multiple opposing directions in which a result could be considered positive. For example, the expression of two genes could be correlated, uncorrelated, or anticorrelated. Both correlation and anticorrelation might be seen as positive results, whereas the null result of no correlation could be subject to publication bias. But suppose there is truly no effect: what does publication bias do in this case? We would expect to see false positives in both directions. Meta-analysis would readily pick up the lack of a consistent direction of the effect, and (if the authors avoid mistakenly inferring population heterogeneity) it is unlikely that correlations in either direction would be falsely canonized as fact. 

Our model assumes that research continues until each claim is either rejected or canonized as fact. In practice, researchers can and do lose interest in certain claims. False claims might generate more conflicting results, or take longer to reach one of the evidentiary thresholds; either mechanism could lead researchers to move on to other problems and leave the claim as unresolved. If this is the case, we might expect that instead of being rejected or canonized as fact, many false claims might simply be abandoned.

Another possible difference between the model and the real world is that we model the evidentiary standards as symmetric, but in practice it may require less certainty to discard a claim as false than it requires to accept the same claim as fact. In this case, the probability of rejecting false claims would be higher than predicted in our model---possibly with only a very small increase in the probability of rejecting true claims. 

The scientific community could also actively respond to the problem of canonizing false claims. One of the most direct ways would be to invest more heavily in the publication of negative results. A number of new journals or collections within journals have been established to specialize in publishing negative results. This includes Elsevier's {\em New Negatives in Plant Science}, {\em PLOS One's} Positively Negative collection, Biomed Central's {\em Journal of Negative Results in Biomedicine}, and many others \cite{nature2016go}. Alternatively, peer reviewed publication may be unnecessary; simply publishing negative results on preprint archives such as the arXiv, bioRxiv, and SocArXiv may make these results sufficiently visible. In either case, we face an incentive problem: if researchers accrue scant credit or reward for their negative findings, there is little reason for them to invest the substantial time needed in taking a negative result from a bench-top disappointment to a formal publication. 

Another possibility---which may already be in play---involves shifting probabilities of publishing negative results. We have shown that if negative results become easier to publish as a claim becomes better established, this can greatly reduce the probability of canonizing false claims. One possibility is that negative results may become easier to publish as they become more surprising to the community, i.e., as researchers become increasingly convinced that a claim is true. Referees and journal editors could make an active effort to value papers of this sort. At present, however, our experience suggests that negative results or even corrections of blatant errors in previous publications rarely land in journals of equal prestige to those that published the original positive studies \cite{matosin2014negativity}.

A final saving grace is that even after false claims are established as facts, science can still self-correct. In this paper, we have assumed for simplicity that claims are independent propositions, but in practice claims are entangled in a web of logical interrelations. When a false claim is canonized as fact, inconsistencies between it and other facts soon begin to accumulate until the field is forced to reevaluate the conflicting facts. Results that resolve these conflicts by disproving accepted facts then take on a special significance and suffer little of the stigma placed upon negative results. Until the scientific community finds more ways to deal with publication bias, this may be an essential corrective to a process that sometimes loses its way.

\section*{Acknowledgments}
The authors thank Luis Amaral, Jacob G.~Foster, Frazer Meacham, Peter Rodgers, Ludo Waltman, and Kevin Zollman for helpful comments and suggestions on the manuscript. This work was supported by the Danish National Research Foundation and by a generous grant to the Metaknowledge Network from the John Templeton Foundation. KG thanks the University of Washington Department of Biology for sabbatical support.

\newpage
\bibliography{facts.bbl}

\begin{thebibliography}{55}
\providecommand{\natexlab}[1]{#1}
\providecommand{\url}[1]{\texttt{#1}}
\expandafter\ifx\csname urlstyle\endcsname\relax
  \providecommand{\doi}[1]{doi: #1}\else
  \providecommand{\doi}{doi: \begingroup \urlstyle{rm}\Url}\fi

\bibitem[Ravetz(1971)]{ravetz1971scientific}
Jerome~R. Ravetz.
\newblock \emph{{Scientific Knowledge and its Social Problems}}.
\newblock British Society for the Philosophy of Science, 1971.
\newblock ISBN 9788578110796.
\newblock \doi{10.1017/CBO9781107415324.004}.

\bibitem[Arbesman(2012)]{arbesman2012halflife}
Samuel Arbesman.
\newblock \emph{{The Half-life of Facts: Why Everything We Know Has an
  Expiration Date}}.
\newblock Penguin, 2012.

\bibitem[Latour(1987)]{latour1987science}
Bruno Latour.
\newblock \emph{{Science in Action: How to Follow Scientists and Engineers
  through Society}}.
\newblock Harvard University Press, 1987.
\newblock ISBN 0-674-79290-4.

\bibitem[Begley and Ellis(2012)]{begley2012drug}
C.~Glenn Begley and Lee~M. Ellis.
\newblock {Drug development: Raise standards for preclinical cancer research}.
\newblock \emph{Nature}, 483\penalty0 (7391):\penalty0 531--3, 2012.
\newblock ISSN 1476-4687.
\newblock \doi{10.1038/483531a}.

\bibitem[{Open Science Collaboration}(2015)]{open2015estimating}
{Open Science Collaboration}.
\newblock {Estimating the reproducibility of psychological science}.
\newblock \emph{Science}, 349\penalty0 (6251):\penalty0 aac4716--aac4716, 2015.
\newblock ISSN 0036-8075.
\newblock \doi{10.1126/science.aac4716}.

\bibitem[Errington et~al.(2014)Errington, Iorns, Gunn, Tan, Lomax, and
  Nosek]{errington2014open}
Timothy~M Errington, Elizabeth Iorns, William Gunn, Fraser~Elisabeth Tan,
  Joelle Lomax, and Brian~A Nosek.
\newblock {An open investigation of the reproducibility of cancer biology
  research}.
\newblock \emph{eLife}, 3:\penalty0 e04333, 2014.
\newblock ISSN 2050-084X.
\newblock \doi{10.7554/eLife.04333}.

\bibitem[Ebrahim et~al.(2014)Ebrahim, Sohani, Montoya, Agarwal, Thorlund,
  Mills, and Ioannidis]{ebrahim2014reanalyses}
Shanil Ebrahim, Zahra~N Sohani, Luis Montoya, Arnav Agarwal, Kristian Thorlund,
  Edward~J Mills, and John P~a Ioannidis.
\newblock {Reanalyses of randomized clinical trial data}.
\newblock \emph{The Journal of the American Medical Association}, 312\penalty0
  (10):\penalty0 1024--32, 2014.
\newblock ISSN 1538-3598.
\newblock \doi{10.1001/jama.2014.9646}.

\bibitem[Chang and Li(2015)]{chang2015economics}
Andrew~C Chang and Phillip Li.
\newblock {Is Economics Research Replicable? Sixty Published Papers from
  Thirteen Journals Say ``Usually Not"}.
\newblock \emph{Finance and Economics Discussion Series}, 083:\penalty0 1--26,
  2015.
\newblock ISSN 19362854.
\newblock \doi{10.17016/FEDS.2015.083}.

\bibitem[Camerer et~al.(2016)Camerer, Dreber, Forsell, Ho, Huber, Kirchler,
  Almenberg, Altmejd, Chan, Holzmeister, Imai, Isaksson, Nave, Pfeiffer, Razen,
  and Wu]{camerer2016evaluating}
Colin~F Camerer, Anna Dreber, Eskil Forsell, Teck-hua Ho, J{\"{u}}rgen Huber,
  Michael Kirchler, Johan Almenberg, Adam Altmejd, Taizan Chan, Felix
  Holzmeister, Taisuke Imai, Siri Isaksson, Gideon Nave, Thomas Pfeiffer,
  Michael Razen, and Hang Wu.
\newblock {Evaluating replicability of laboratory experiments in economics}.
\newblock \emph{Science}, 351\penalty0 (6280):\penalty0 1433--1436, 2016.
\newblock ISSN 0036-8075.
\newblock \doi{10.1126/science.aaf0918}.

\bibitem[Baker(2016)]{Baker2016}
Monya Baker.
\newblock 1,500 scientists lift the lid on reproducibility.
\newblock \emph{Nature}, 533\penalty0 (7604):\penalty0 452--454, 2016.

\bibitem[Ioannidis(2005)]{Ioannidis2005}
John P~A Ioannidis.
\newblock {Why Most Published Research Findings Are False}.
\newblock \emph{PLoS Medicine}, 2\penalty0 (8), 2005.
\newblock ISSN 15491277.
\newblock \doi{10.1371/journal.pmed.0020124}.

\bibitem[Higginson and Munaf{\`{o}}(2016)]{higginson2016current}
Andrew~D. Higginson and Marcus~R. Munaf{\`{o}}.
\newblock {Current Incentives for Scientists Lead to Underpowered Studies with
  Erroneous Conclusions}.
\newblock \emph{PLOS Biology}, 14\penalty0 (11):\penalty0 e2000995, 2016.
\newblock ISSN 1545-7885.
\newblock \doi{10.1371/journal.pbio.2000995}.

\bibitem[Bernstein et~al.(2001)Bernstein, Caudy, Hammond, and
  Hannon]{bernstein2001role}
Emily Bernstein, Amy~A Caudy, Scott~M Hammond, and Gregory~J Hannon.
\newblock {Role for a bidentate ribonuclease in the initiation step of RNA
  interference}.
\newblock \emph{Nature}, 409\penalty0 (6818):\penalty0 363--366, 2001.
\newblock ISSN 0028-0836.
\newblock \doi{10.1038/35053110}.

\bibitem[Jaskiewicz and Filipowicz(2008)]{jaskiewicz2008role}
Lukasz Jaskiewicz and Witold Filipowicz.
\newblock {Role of Dicer in posttranscriptional RNA silencing}.
\newblock \emph{Current Topics in Microbiology and Immunology}, 320:\penalty0
  77--97, 2008.

\bibitem[McElreath and Smaldino(2015)]{mcelreath2015replication}
Richard McElreath and Paul~E. Smaldino.
\newblock {Replication, Communication, and the Population Dynamics of
  Scientific Discovery}.
\newblock \emph{PLoS ONE}, 10\penalty0 (8), 2015.
\newblock ISSN 19326203.
\newblock \doi{10.1371/journal.pone.0136088}.

\bibitem[Sterling(1959)]{sterling1959publication}
Theodore~D Sterling.
\newblock {Publication Decisions and Their Possible Effects on Inferences Drawn
  from Tests of Significance --Or Vice Versa}.
\newblock \emph{Journal of the American Statistical Association}, 54\penalty0
  (285):\penalty0 30--34, 1959.

\bibitem[Rosenthal(1979)]{rosenthal1979file}
Robert Rosenthal.
\newblock {The ``File Drawer Problem" and Tolerance for Null Results}.
\newblock \emph{Psychological Bulletin}, 86\penalty0 (3):\penalty0 638--641,
  1979.
\newblock ISSN 0033-2909.
\newblock \doi{10.1037/0033-2909.86.3.638}.

\bibitem[Newcombe(1987)]{newcombe1987towards}
Robert~G Newcombe.
\newblock {Towards a reduction in publication bias}.
\newblock \emph{British Medical Journal}, 295\penalty0 (6599):\penalty0
  656--659, 1987.
\newblock ISSN 0959-8138.
\newblock \doi{10.1136/bmj.295.6599.656}.

\bibitem[Begg and Berlin(1988)]{begg1988publication}
Colin~B Begg and Jesse~A Berlin.
\newblock {Publication Bias: A Problem in Interpreting Medical Data}.
\newblock \emph{Source Journal of the Royal Statistical Society. Series A
  (Statistics in Society) J. R. Statist. Soc. A}, 151\penalty0 (3):\penalty0
  419--463, 1988.
\newblock ISSN 09641998.
\newblock \doi{10.2307/2982993}.

\bibitem[Dickersin(1990)]{dickersin1990existence}
Kay Dickersin.
\newblock {The Existence of Publication Bias and Risk Factors for its
  Occurrence}.
\newblock \emph{The Journal of the American Medical Association}, \penalty0
  (10):\penalty0 1385, 1990.

\bibitem[Easterbrook et~al.(1991)Easterbrook, Berlin, Gopalan, and
  Matthews]{easterbrook1991publication}
Philippa~J Easterbrook, Jesse~A Berlin, Ramana Gopalan, and David~R Matthews.
\newblock {Publication bias in clinical research}.
\newblock \emph{The Lancet}, 337\penalty0 (8746):\penalty0 867--872, 1991.
\newblock ISSN 01406736.
\newblock \doi{10.1016/0140-6736(91)90201-Y}.

\bibitem[Song et~al.(2000)Song, Eastwood, Gilbody, Duley, and
  Sutton]{song2000publication}
F.~Song, A.~J. Eastwood, S.~Gilbody, L.~Duley, and A.~J. Sutton.
\newblock {Publication and related biases}.
\newblock \emph{Health Technology Assessment}, 4\penalty0 (10), 2000.
\newblock ISSN 13665278.
\newblock \doi{Health Technology Assessment}.

\bibitem[Olson et~al.(2002)Olson, Rennie, Cook, Dickersin, Flanagin, Hogan,
  Zhu, Reiling, and Pace]{olson2002publication}
Carin~M Olson, Drummond Rennie, Deborah Cook, Kay Dickersin, Annette Flanagin,
  Joseph~W Hogan, Qi~Zhu, Jennifer Reiling, and Brian Pace.
\newblock {Publication Bias in Editorial Decision Making}.
\newblock \emph{The Journal of the American Medical Association}, 287\penalty0
  (21):\penalty0 2825--2828, 2002.
\newblock ISSN 00987484.
\newblock \doi{10.1001/jama.287.21.2825}.

\bibitem[Chan and Altman(2005)]{chan2005identifying}
An-Wen Chan and Douglas~G Altman.
\newblock {Identifying outcome reporting bias in randomised trials on PubMed:
  review of publications and survey of authors}.
\newblock \emph{BMJ}, 330\penalty0 (7494):\penalty0 753, 2005.
\newblock ISSN 1756-1833.
\newblock \doi{10.1136/bmj.38356.424606.8F}.

\bibitem[Franco et~al.(2014)Franco, Malhotra, and
  Simonovits]{franco2014publication}
Annie Franco, Neil Malhotra, and Gabor Simonovits.
\newblock {Publication bias in the social sciences: Unlocking the file drawer}.
\newblock \emph{Science}, 345\penalty0 (6203):\penalty0 1502--1505, 2014.

\bibitem[Turner et~al.(2008)Turner, Matthews, Linardatos, Tell, and
  Rosenthal]{turner2008selective}
Erick~H. Turner, Annette~M. Matthews, Eftihia Linardatos, Robert~A. Tell, and
  Robert Rosenthal.
\newblock {Selective Publication of Antidepressant Trials and Its Influence on
  Apparent Efficacy}.
\newblock \emph{The New England Journal of Medicine}, 358\penalty0
  (3):\penalty0 252--260, 2008.
\newblock ISSN 0028-4793.
\newblock \doi{10.1056/NEJMsa065779}.

\bibitem[Csada et~al.(1996)Csada, James, and Espie]{csada1996file}
Ryan~D Csada, Paul~C James, and Richard H~M Espie.
\newblock {The "File Drawer Problem" of Non-Significant Resluts: Does It Apply
  to Biologica Research?}
\newblock \emph{Oikos}, 76\penalty0 (3):\penalty0 591--593, 1996.

\bibitem[Fanelli(2012)]{fanelli2011negative}
Daniele Fanelli.
\newblock {Negative results are disappearing from most disciplines and
  countries}.
\newblock \emph{Scientometrics}, 90\penalty0 (3):\penalty0 891--904, 2012.
\newblock ISSN 01389130.
\newblock \doi{10.1007/s11192-011-0494-7}.

\bibitem[Rzhetsky et~al.(2006)Rzhetsky, Iossifov, Loh, and
  White]{rzhetsky2006microparadigms}
Andrey Rzhetsky, Ivan Iossifov, Ji~Meng Loh, and Kevin~P White.
\newblock {Microparadigms: Chains of collective reasoning in publications about
  molecular interactions}.
\newblock \emph{Proceedings of the National Academy of Sciences of the United
  States of America}, 103\penalty0 (13):\penalty0 4940--5, 2006.
\newblock ISSN 0027-8424.
\newblock \doi{10.1073/pnas.0600591103}.

\bibitem[Norris(1998)]{norris1998markov}
James~R. Norris.
\newblock \emph{{Markov Chains}}.
\newblock Cambridge University Press, 1998.

\bibitem[Head et~al.(2015)Head, Holman, Lanfear, Kahn, and
  Jennions]{head2015extent}
Megan~L. Head, Luke Holman, Rob Lanfear, Andrew~T. Kahn, and Michael~D.
  Jennions.
\newblock {The Extent and Consequences of P-Hacking in Science}.
\newblock \emph{PLoS Biology}, 13\penalty0 (3), 2015.
\newblock ISSN 15457885.
\newblock \doi{10.1371/journal.pbio.1002106}.

\bibitem[{Le Noury} et~al.(2015){Le Noury}, Nardo, Healy, Jureidini, Raven,
  Tufanaru, and Abi-Jaoude]{le2015restoring}
Joanna {Le Noury}, John~M Nardo, David Healy, Jon Jureidini, Melissa Raven,
  Catalin Tufanaru, and Elia Abi-Jaoude.
\newblock {Restoring Study 329: efficacy and harms of paroxetine and imipramine
  in treatment of major depression in adolescence.}
\newblock \emph{BMJ (Clinical research ed.)}, 351\penalty0 (7):\penalty0 h4320,
  2015.
\newblock ISSN 1756-1833.
\newblock \doi{10.1136/bmj.h4320}.

\bibitem[Tannock(1996)]{tannock1996false}
Ian~F. Tannock.
\newblock {False-Positive Results in Clinical Trials: Multiple Significance
  Tests and the Problem of Unreported Comparisons}.
\newblock \emph{Journal of the National Cancer Institute}, 88\penalty0
  (3/4):\penalty0 206--207, 1996.

\bibitem[Smith and Ebrahim(2002)]{smith2002data}
George~Davey Smith and Shah Ebrahim.
\newblock {Data dredging, bias, or confounding}.
\newblock \emph{BMJ (Clinical research ed.)}, 325\penalty0 (7378):\penalty0
  1437--1438, 2002.
\newblock ISSN 09598138.
\newblock \doi{10.1136/bmj.325.7378.1437}.

\bibitem[Kerr(1998)]{kerr1998harking}
Norbert~L Kerr.
\newblock {HARKing: Hypothesizing After the Results are Known}.
\newblock \emph{Personality and Social Psychology Review}, 2\penalty0
  (3):\penalty0 196--217, 1998.

\bibitem[Gelman and Loken(2014)]{gelman2014statistical}
Andrew Gelman and Eric Loken.
\newblock {The Statistical Crisis in Science}.
\newblock \emph{American Scientist}, 102\penalty0 (6):\penalty0 460--465, 2014.

\bibitem[Pocock et~al.(1987)Pocock, Hughes, and Lee]{pocock1987statistical}
Stuart~J Pocock, Michael~D Hughes, and Robert~J Lee.
\newblock {Statistical Problems in the Reporting of Clinical Trials}.
\newblock \emph{The New England Journal of Medicine}, 317\penalty0
  (7):\penalty0 426--32, 1987.
\newblock ISSN 0028-4793.
\newblock \doi{10.1056/NEJM198708133170706}.

\bibitem[Eklund et~al.(2016)Eklund, Nichols, and Knutsson]{eklund2016cluster}
Anders Eklund, Thomas~E. Nichols, and Hans Knutsson.
\newblock {Cluster failure: Why fMRI inferences for spatial extent have
  inflated false-positive rates}.
\newblock \emph{Proceedings of the National Academy of Sciences}, page
  201602413, 2016.
\newblock ISSN 0027-8424.
\newblock \doi{10.1073/pnas.1602413113}.

\bibitem[Kasenda et~al.(2016)Kasenda, von Elm, You, Bl{\"u}mle, Tomonaga,
  Saccilotto, Amstutz, Bengough, Meerpohl, Stegert,
  et~al.]{kasenda2016agreements}
Benjamin Kasenda, Erik von Elm, John~J You, Anette Bl{\"u}mle, Yuki Tomonaga,
  Ramon Saccilotto, Alain Amstutz, Theresa Bengough, Joerg~J Meerpohl, Mihaela
  Stegert, et~al.
\newblock Agreements between industry and academia on publication rights: A
  retrospective study of protocols and publications of randomized clinical
  trials.
\newblock \emph{PLoS Med}, 13\penalty0 (6):\penalty0 e1002046, 2016.

\bibitem[Silvertown and McConway(1997)]{silvertown1997does}
Jonathan Silvertown and Kevin~J. McConway.
\newblock {Does ``Publication Bias" Lead to Biased Science?}
\newblock \emph{Oikos}, 79\penalty0 (1):\penalty0 167--168, 1997.

\bibitem[Ioannidis and Trikalinos(2005)]{ioannidis2005early}
John P~A Ioannidis and Thomas~A. Trikalinos.
\newblock {Early extreme contradictory estimates may appear in published
  research: The Proteus phenomenon in molecular genetics research and
  randomized trials}.
\newblock \emph{Journal of Clinical Epidemiology}, 58\penalty0 (6):\penalty0
  543--549, 2005.
\newblock ISSN 08954356.
\newblock \doi{10.1016/j.jclinepi.2004.10.019}.

\bibitem[de~Winter and Happee(2013)]{de2013selective}
Joost de~Winter and Riender Happee.
\newblock {Why Selective Publication of Statistically Significant Results Can
  Be Effective}.
\newblock \emph{PLoS ONE}, 8\penalty0 (6), 2013.
\newblock ISSN 19326203.
\newblock \doi{10.1371/journal.pone.0066463}.

\bibitem[van Assen et~al.(2014)van Assen, van Aert, Nuijten, and
  Wicherts]{van2014publishing}
Marcel A L~M van Assen, Robbie C~M van Aert, Mich{\`{e}}le~B Nuijten, and
  Jelte~M Wicherts.
\newblock {Why Publishing Everything Is More Effective than Selective
  Publishing of Statistically Significant Results}.
\newblock \emph{PLoS ONE}, 9\penalty0 (1), 2014.
\newblock ISSN 1932-6203.
\newblock \doi{10.1371/journal.pone.0084896}.

\bibitem[Poulin(2000)]{poulin2000manipulation}
R~Poulin.
\newblock {Manipulation of host behaviour by parasites: a weakening paradigm?}
\newblock \emph{Proceedings of the Royal Society of London B: Biological
  Sciences}, 267\penalty0 (1445):\penalty0 787--792, 2000.

\bibitem[Simmons et~al.(1999)Simmons, Tomkins, Kotiaho, and
  Hunt]{simmons1999fluctuating}
Leigh~W Simmons, J~L Tomkins, J~S Kotiaho, and J~Hunt.
\newblock {Fluctuating paradigm}.
\newblock \emph{Proceedings of the Royal Society of London B: Biological
  Sciences}, 266\penalty0 (1419):\penalty0 593--595, 1999.
\newblock ISSN 0962-8452.
\newblock \doi{10.1098/rspb.1999.0677}.

\bibitem[Palmer(2000)]{palmer2000quasireplication}
A.~Richard Palmer.
\newblock {Quasi-Replication and the Contract of Error: Lessons from Sex
  Ratios, Heritabilities and Fluctuating Asymmetry}.
\newblock \emph{Annual Review of Ecology and Systematics}, 31\penalty0
  (1):\penalty0 441--480, 2000.

\bibitem[Jennions and M{\o}ller(2002)]{jennions2002relationships}
Michael~D Jennions and Anders~P M{\o}ller.
\newblock {Relationships fade with time: a meta-analysis of temporal trends in
  publication in ecology and evolution}.
\newblock \emph{Proceedings of the Royal Society of London B: Biological
  Sciences}, 269\penalty0 (1486):\penalty0 43--48, 2002.
\newblock ISSN 0962-8452.
\newblock \doi{10.1098/rspb.2001.1832}.

\bibitem[Knight(2003)]{knight2003negative}
Jonathan Knight.
\newblock {Negative results: Null and void}.
\newblock \emph{Nature}, 422\penalty0 (6932):\penalty0 554--555, 2003.
\newblock ISSN 0028-0836.
\newblock \doi{10.1038/422554a}.

\bibitem[Nelson et~al.(2012)Nelson, Simmons, and Simonsohn]{nelson2012let}
Leif~D Nelson, Joseph~P Simmons, and Uri Simonsohn.
\newblock {Let's Publish Fewer Papers}.
\newblock \emph{Psychological Inquiry}, 233\penalty0 (23):\penalty0 291--293,
  2012.
\newblock ISSN 1532-7965.
\newblock \doi{10.1080/1047840X.2012.705245}.

\bibitem[Savage(1954)]{savage1954foundations}
Leonard~J. Savage.
\newblock \emph{{The foundations of statistics}}.
\newblock John Wiley {\&} Sons, 1954.

\bibitem[Good(1967)]{good1967principle}
I.~J. Good.
\newblock {On the Principle of Total Evidence}.
\newblock \emph{British Journal for the Philosophy of Science}, 17\penalty0
  (4):\penalty0 319--321, 1967.

\bibitem[Ramsey(1990)]{ramsey1990weight}
F.~P. Ramsey.
\newblock {Weight or the Value of Knowledge}.
\newblock \emph{The British Journal for the Philosophy of Science}, 41\penalty0
  (1):\penalty0 1--4, 1990.

\bibitem[Dickersin et~al.(1992)Dickersin, Min, and
  Meinert]{dickersin1992factors}
Kay Dickersin, Yuan-I Min, and Curtis~L Meinert.
\newblock {Factors influencing publication of research results. Follow-up of
  applications submitted to two institutional review boards}.
\newblock \emph{The Journal of the American Medical Association}, 267\penalty0
  (3):\penalty0 374--8, 1992.

\bibitem[Editors(2016)]{nature2016go}
Editors.
\newblock {Go forth and replicate!}
\newblock \emph{Nature}, 536:\penalty0 373, 2016.

\bibitem[Matosin et~al.(2014)Matosin, Frank, Engel, Lum, and
  Newell]{matosin2014negativity}
Natalie Matosin, Elisabeth Frank, Martin Engel, Jeremy~S Lum, and Kelly~a
  Newell.
\newblock {Negativity towards negative results: a discussion of the disconnect
  between scientific worth and scientific culture}.
\newblock \emph{Disease Models {\&} Mechanisms}, 7\penalty0 (2):\penalty0
  171--3, 2014.
\newblock ISSN 1754-8411.
\newblock \doi{10.1242/dmm.015123}.

\end{thebibliography}

\end{document}